\documentclass[aps, prl, twocolumn, superscriptaddress,preprintnumbers]{revtex4-2}
\usepackage{bm, amsmath, amsfonts, amssymb, ascmac, mathtools, braket}
\usepackage{times}
\usepackage{multirow}
\usepackage{graphicx}
\usepackage{float, color, xcolor}

\usepackage[whole]{bxcjkjatype} 
\usepackage{subfigure} 
\usepackage{amscd} 

\usepackage{bbm}
\usepackage{tabularx}

\usepackage{comment}

\usepackage[
pagebackref=false,
colorlinks=true,
linkcolor=blue,
urlcolor=blue,
filecolor=black,
citecolor=red,
pdfstartview=FitV,
pdftitle={},
pdfauthor={},
pdfsubject={},
pdfkeywords={},
pdfpagemode=None,
bookmarksopen=true
]{hyperref}

\newcommand{\ii}{\mathrm{i}}
\newcommand{\Tr}{\operatorname{Tr}}
\newcommand{\dif}{\mathrm{d}}

\newcommand{\SIBC}{\mathrm{SIBC}}
\newcommand{\sgn}{\operatorname{sgn}}

\newcommand{\modified}[2]{#1}{}

\newcommand{\whatistitle}{Subspace-Protected Topological Phases and Bulk-Boundary Correspondence}

\begin{document}

\title{\whatistitle}

\author{Kenji Shimomura}
\email{kenji.shimomura@yukawa.kyoto-u.ac.jp}
\affiliation{Center for Gravitational Physics and Quantum Information, Yukawa Institute for Theoretical Physics, Kyoto University, Kyoto 606-8502, Japan}

\author{Ryo Takami}
\email{ryo.takami@yukawa.kyoto-u.ac.jp}
\affiliation{Center for Gravitational Physics and Quantum Information, Yukawa Institute for Theoretical Physics, Kyoto University, Kyoto 606-8502, Japan}

\author{Daichi Nakamura}
\email{daichi.nakamura@issp.u-tokyo.ac.jp}
\affiliation{Institute for Solid State Physics, University of Tokyo, Kashiwa, Chiba 277-8581, Japan}

\author{Masatoshi Sato}
\email{msato@yukawa.kyoto-u.ac.jp}
\affiliation{Center for Gravitational Physics and Quantum Information, Yukawa Institute for Theoretical Physics, Kyoto University, Kyoto 606-8502, Japan}

\date{\today}

\begin{abstract}
    While tremendous research has revealed that symmetry enriches topological phases of matter, more general principles that protect topological phases have yet to be explored. In this Letter, we elucidate the roles of subspaces in free-fermionic topological phases. A subspace property for Hamiltonians enables us to define new topological invariants. It results in peculiar topological boundary phenomena, i.e., the emergence of an unpaired zero mode or zero-winding skin modes, characterizing subspace-protected topological phases. We establish and demonstrate the bulk-boundary correspondence in subspace-protected topological phases. We further discuss the interplay of the subspace property and internal symmetries. Toward application, we also propose possible platforms possessing the subspace property.
\end{abstract}

\maketitle

Symmetry is a fundamental principle to characterize and classify topological phases of matter \cite{Wen2017}.
In particular, various types of symmetry provide free-fermion systems with rich topological structure, as in topological insulators and superconductors \cite{Moore2010,Hasan2010,Qi2011,Chiu2016,Sato2017}.
A unitary symmetry resolves the Hilbert space into sectors, in which topological phases are individually classified on the basis of internal symmetries \cite{Schnyder2008,Schnyder2009,Kitaev2009,Ryu2010,Teo2010,Freed2013} or crystalline symmetries \cite{Fu2011,Hsieh2012,Slager2013,Chiu2013,Morimoto2013,Shiozaki2014,Kruthoff2017,Po2017,Bradlyn2017}.
A significant characteristic of topological phases is the bulk-boundary correspondence: the nontrivial topological phases in gapped bulk involves the presence of gapless boundary modes \cite{Hatsugai1993,Hatsugai1993b,Elbau2002,Mong2011,Essin2011,Graf2013,Witten2016,Prodan2016}.
Nontrivial topological insulators protected by time-reversal symmetry (TRS) exhibit helical gapless modes as the Kramers pairs \cite{Ryu2002,Bernevig2005,Kane2005,Kane2005b,Bernevig2006,Fu2007,Fu2007b,Moore2007,Teo2008,Roy2009}, and nontrivial topological superconductors can host Majorana zero modes reflecting particle-hole symmetry \cite{Read2000,Ivanov2001,Kitaev2001,Sato2003,Fu2008,Volovik2009,Qi2009,Sato2009,Sau2010,Lutchyn2010,Alicea2010,Oreg2010,Tanaka2012}.

The scheme of topological phases is extended into non-Hermitian systems with symmetries and gap structures enlarged from Hermitian cases \cite{Rudner2009,Sato2012,Hu2011,Esaki2011,Schomerus2013,Longhi2015,Lee2016,Leykam2017,Xu2017,Shen2018,Kozii2024,Takata2018,Gong2018,Kawabata2019,Yao2018,Yao2018b,Kunst2018,McDonald2018,Lee2019,Liu2019,Lee2019b,Kawabata2019b,Zhou2019,Herviou2019,Zirnstein2021,Borgnia2020,Kawabata2019c,Yokomizo2019,Okuma2019,Lee2019c,Wanjura2020,Zhang2020,Okuma2020,Sone2020,Yi2020,Kawabata2020,Terrier2020,Bessho2021,Denner2021,Okugawa2020,Kawabata2020b,Kawabata2021,Zhang2022,Sun2021,Delplace2021,Franca2022,Yoshida2022,Nakamura2024,Wang2024,Ma2024,Schindler2023,Nakai2024,Nakamura2023,Denner2023,Hamanaka2024,Nakamura2024b,Shiozaki2024,Ashida2020,Bergholtz2021,Okuma2023,Lin2023,Zhu2024}.
Above all, point-gapped topological phases are intrinsic in non-Hermitian bulk systems \cite{Kawabata2019b} and can cause anomalous boundary behaviors such as non-Hermitian skin effects \cite{Lee2016,Yao2018,Okuma2020,Zhang2020} or the emergence of exceptional boundary states \cite{Terrier2020,Denner2021,Nakamura2024}.
Symmetry also enriches the point-gapped topological phases; 
for instance, $\mathbb{Z}_2$ skin effects occur protected by transpose-type time-reversal symmetry (TRS$^\dag$) \cite{Okuma2020}.

However, several topological phenomena in free-fermion systems challenge explanation beyond conventional notion of symmetry.
For instance, recent studies have observed topological zero modes protected by a so-called sub-symmetry in some Hermitian systems with broken sublattice symmetry \cite{Jangjan2022,Wang2023,Liu2023,Verma2024,Verma2024b,Kang2024}.
In non-Hermitian cases, trivial point-gapped systems have been reported to host skin modes even with no symmetry \cite{Guo2023,Zhou2025}.
Thus, exploring general principles protecting topological phenomena is an elusive but urgent issue.

\begin{figure}
    \centering
    \includegraphics[width=0.9\linewidth]{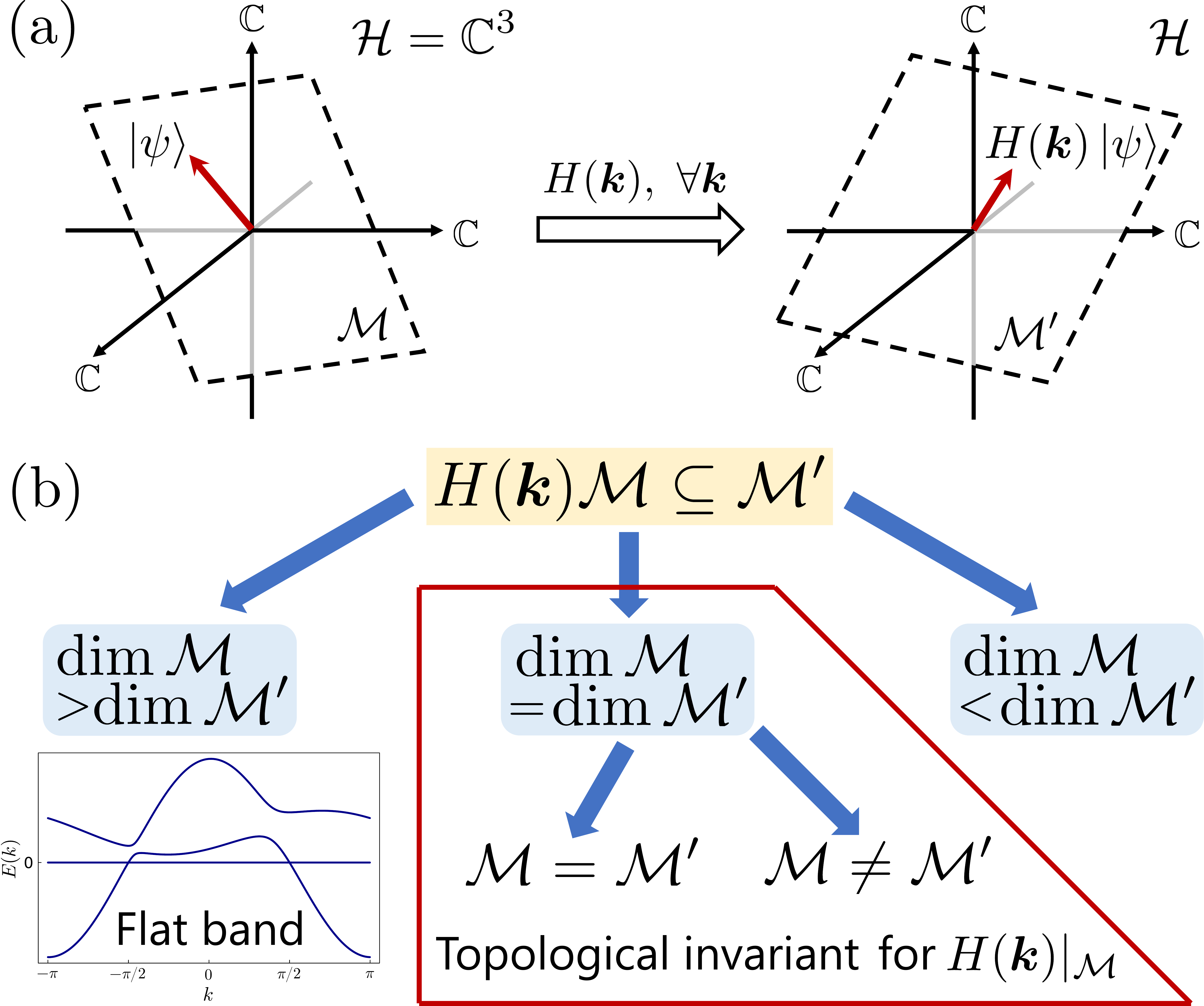}
    \caption{(a) Schematic figure of the subspace property with two-dimensional subspaces $\mathcal{M},\mathcal{M}'$ in the three-dimensional Hilbert space $\mathcal{H}$. 
    The subspace property is regarded as a geometric condition on the Hilbert space.
    (b) Classification of relations between $\mathcal{M}$ and $\mathcal{M}'$. In the case of $\dim\mathcal{M}=\dim\mathcal{M}'$, the subspace-protected topological invariant is well-defined.}
    \label{fig:schematic_figure}
\end{figure}

In this Letter, we reveal that subspaces of Hilbert spaces 
serve as a principle to introduce a novel type of free-fermion topological phases in both Hermitian and non-Hermitian systems beyond the conventional notion of symmetry.
Specifically, we consider the following condition, which we call the \textit{subspace property}, for a Bloch Hamiltonian $H(\boldsymbol{k})$ acting on a Hilbert space $\mathcal{H}$ of internal degrees of freedom,
\begin{align}\label{eq:subspace_property}
    H(\boldsymbol{k})\mathcal{M}
    \subseteq \mathcal{M}',
\end{align}
where $\mathcal{M}$ and $\mathcal{M}'$ are subspaces of $\mathcal{H}$ independent of the wave vector $\boldsymbol{k}$, as illustrated in Fig.~\ref{fig:schematic_figure} (a).
Note that the subspace property is not reduced to group-theoretical symmetry operations.
We propose the concept of \modified{gapped}{} topological phases protected by the subspace property \eqref{eq:subspace_property}, dubbed \textit{subspace-protected topological phases}, and establish the bulk-boundary correspondence in the phases.
Furthermore, we clarify the interplay of subspace-protected topological phases and internal symmetries.
Toward application, we provide searching guidelines for physical platforms possessing the subspace property.

\textit{Consequences of the subspace property}.---
Prior to the investigation of subspace-protected topological phases, let us explain a physical interpretation of the subspace property.
Although the subspace property in Eq.~\eqref{eq:subspace_property} is apparently mathematical, it is equivalent to a selection rule,
\begin{align}\label{eq:selection_rule}
    \bra{\psi'}H(\boldsymbol{k})\ket{\psi}
    = 0,
    \quad
    \ket{\psi}\in\mathcal{M},\ \ket{\psi'}\in\mathcal{M}^{\prime\perp}.
\end{align}
As well as familiar selection rules rooted in symmetry, Eq.~\eqref{eq:selection_rule} may prohibit particular backscattering processes, allowing boundary modes to stabilize against perturbations.
Therefore, we should have a topological invariant protected by the subspace property to clarify the stable boundary modes.
Later, we will rigorously justify this intuition.

To define topological phases, we identify a condition that systems with the subspace property have gapped spectra \modified{in the point-gap sense \cite{Kawabata2019b}}{},
\modified{
\begin{align}\label{eq:point-gap}
    \forall\boldsymbol{k},
    \quad
    \det H(\boldsymbol{k})
    \neq 0,
\end{align}
indicating the absence of zero-energy modes in the bulk, where $H(\boldsymbol{k})$ may or may not be Hermitian.}{}
For this purpose, we classify the subspace property according to the relation between the dimensions of $\mathcal{M}$ and $\mathcal{M}'$ as shown in Fig.~\ref{fig:schematic_figure} (b).

If $m\coloneqq\dim\mathcal{M}>\dim\mathcal{M}'$, then $H(\boldsymbol{k})$ hosts at least a flat band at zero energy in the bulk spectrum:
For linearly independent states $\ket{e_1},\ldots,\ket{e_m}\in\mathcal{M}\setminus\{0\}$, $m$ states $H(\boldsymbol{k})\ket{e_1},\ldots, H(\boldsymbol{k})\ket{e_m}\in\mathcal{M}'$ are linearly dependent, and thus, there is a particular superposition $H(\boldsymbol{k})(c_1(\boldsymbol{k})\ket{e_1}+\cdots+c_m(\boldsymbol{k})\ket{e_m})
=0$ for every $\boldsymbol{k}$, giving a flat band.
This flat band \modified{violates the point-gap condition \eqref{eq:point-gap} and thus}{} hinders defining any topological invariant to detect zero modes, so we exclude this case from our consideration hereafter.

The case of $\dim\mathcal{M}<\dim\mathcal{M}'$ includes the situation of $\mathcal{M}'=\mathcal{H}$, for which the subspace property $H(\boldsymbol{k})\mathcal{M}\subseteq\mathcal{H}$ always holds for generic Hamiltonian $H(\boldsymbol{k})$.
Hence, \modified{there is no subspace-protected topological invariant commonly applicable to arbitrary $\mathcal{M}$ and $\mathcal{M}'$. We could obtain some special results for particular $\mathcal{M}$ and $\mathcal{M}'$, but we leave it for the future.}{the subspace property may bring nothing.}

Thus, we focus on the case of $m\coloneqq\dim\mathcal{M}=\dim\mathcal{M}'$.
This case is further divided into two cases, $\mathcal{M}=\mathcal{M}'$ and $\mathcal{M}\neq\mathcal{M}'$.
A difference between these cases appears in the stability of the subspace property with respect to energy shifts.
If $\mathcal{M}=\mathcal{M}'$, the subspace property $H(\boldsymbol{k})\mathcal{M}\subseteq\mathcal{M}$ for $H(\boldsymbol{k})$ leads to that for the shifted Hamiltonian $(H(\boldsymbol{k})-c)$, as $(H(\boldsymbol{k})-c)\mathcal{M}\subseteq\mathcal{M}$.
By contrast, if $\mathcal{M}\neq\mathcal{M}'$, the energy shift $H(\boldsymbol{k})\mapsto H(\boldsymbol{k})-c$ breaks the subspace property due to $c\mathcal{M}\neq\mathcal{M}'$.
As we will see below, this different behavior results in distinct topological phenomena.


\textit{Subspace-protected topological invariant and bulk-boundary correspondence}.---
To introduce topological invariant under the subspace property, we assume both the presence of \modified{the}{} point gap \modified{in Eq.~\eqref{eq:point-gap}}{in $H(\boldsymbol{k})$ at zero energy \cite{Kawabata2019b},
\begin{align}\label{eq:point-gap}
    \forall\boldsymbol{k},
    \quad
    \det H(\boldsymbol{k})
    \neq 0,
\end{align}}{}
and the subspace property \eqref{eq:subspace_property} with $\dim\mathcal{M}=\dim\mathcal{M}'$.
Here, we do not suppose $H(\boldsymbol{k})$ is Hermitian; The discussion below is valid for both Hermitian and non-Hermitian systems.

Our strategy to construct the topological invariant is utilizing the restriction $H(\boldsymbol{k})|_{\mathcal{M}}\colon\mathcal{M}\to\mathcal{M}'$ of $H(\boldsymbol{k})$ onto the subspace $\mathcal{M}$.
\modified{Given orthonormal bases $\ket{e_1},\ldots,\ket{e_m}$ of $\mathcal{M}$ and $\ket{f_1},\ldots,\ket{f_m}$ of $\mathcal{M'}$, a matrix representation of $H(\boldsymbol{k})|_{\mathcal{M}}$ is explicitly given by an $m\times m$ matrix $\sum_{i,i'=1}^m \ket{f_{i'}}\braket{f_{i'}|H(\boldsymbol{k})|e_i}\bra{e_i}$.}{}
We can show that $H(\boldsymbol{k})|_{\mathcal{M}}$ 
is also point-gapped:
From Eq.~\eqref{eq:point-gap}, $H(\boldsymbol{k})$ is invertible, then from $\dim\mathcal{M}=\dim\mathcal{M}'$, 
$H(\boldsymbol{k})|_{\mathcal{M}}$ is also invertible.
Thus, $H(\boldsymbol{k})|_{\mathcal{M}}$ is point-gapped at zero energy, 
{\it i.e.} $\det H(\boldsymbol{k})|_{\mathcal{M}}\neq 0$.

Therefore, we can introduce an additional point-gap topological invariant for $H(\boldsymbol{k})|_{\mathcal{M}}$.
Specifically, in the absence of any symmetry, we have the winding number for odd spatial dimension $d=2n+1$ as
\begin{align}\label{eq:winding}
    w^\mathcal{M}
    \coloneqq \frac{(-1)^n n!}{(2\pi\ii)^{n+1}(2n+1)!}\int_{\text{BZ}}\Tr_\mathcal{M}\left[(H|_{\mathcal{M}}^{-1}\dif H|_{\mathcal{M}})^{2n+1}\right],
\end{align}
where $\text{BZ}$ is the $d$-dimensional Brillouin zone \cite{Kawabata2019b}.
The quantity $w^\mathcal{M}$ is quantized in an integer and invariant under continuous deformation of $H(\boldsymbol{k})$ preserving the point gap and the subspace property with fixed $\mathcal{M}$ and $\mathcal{M}'$. Thus, $w^\mathcal{M}$ is a \textit{subspace-protected topological invariant}.

The subspace-protected topological invariant in the bulk gives rise to characteristic boundary behaviors under open boundary conditions.
To elucidate this bulk-boundary correspondence, we consider a semi-infinite system on $X=\mathbb{Z}_{\ge 0}\times\mathbb{Z}^{d-1}$ with a single boundary in one direction.

Let $H^{\SIBC}\curvearrowright\mathcal{H}\otimes\ell^2(X)$ be the real space Hamiltonian under the semi-infinite boundary condition (SIBC) corresponding to the Bloch Hamiltonian $H(\boldsymbol{k})\curvearrowright\mathcal{H}$, where $\ell^2(X)$ is the Hilbert space of square-summable functions on $X$\footnote{Hereafter we use a symbol $A\curvearrowright\mathcal{K}$ to denote that $A$ is an operator mapping a space $\mathcal{K}$ to itself, $A\colon\mathcal{K}\to\mathcal{K}$. In the End Matter, we summarize notations used in this paper.}.
In the same way, we make the real space SIBC Hamiltonian $H|_{\mathcal{M}}^{\SIBC}\colon\mathcal{M}\otimes\ell^2(X)\to\mathcal{M}'\otimes\ell^2(X)$ corresponding to the restricted Hamiltonian $H(\boldsymbol{k})|_{\mathcal{M}}$. 
As shown below, $H|_{\mathcal{M}}^{\SIBC}$ can host zero eigenvalues, the multiplicity of which is bounded by the subspace-protected topological invariant $w^\mathcal{M}$.
To see this, we introduce the Hermitian Hamiltonian
\begin{align}\label{eq:doubled}
    \tilde{H}(\boldsymbol{k})
    \coloneqq \begin{pmatrix}
        0 & H(\boldsymbol{k})|_{\mathcal{M}}^\dagger \\
        H(\boldsymbol{k})|_{\mathcal{M}} & 0
    \end{pmatrix}
    \curvearrowright\mathcal{M}\oplus\mathcal{M}'
\end{align}
and the corresponding SIBC Hamiltonian $\tilde{H}^{\SIBC}\curvearrowright(\mathcal{M}\oplus\mathcal{M}')\otimes\ell^2(X)$.
Because $\tilde{H}(\boldsymbol{k})$ respects the sublattice symmetry (SLS), $\Gamma\tilde{H}(\boldsymbol{k})\Gamma^{-1}=-\tilde{H}(\boldsymbol{k})$, with $\Gamma=\begin{pmatrix} I_\mathcal{M} & 0 \\ 0 & -I_{\mathcal{M}'} \end{pmatrix}$ and $w^\mathcal{M}$ is the topological invariant for $\tilde{H}(\boldsymbol{k})$ protected by the SLS, the index theorem tells \cite{Essin2011,Sato2011,Prodan2016} 
\begin{align}
    N_+ - N_-
    = w^{\mathcal{M}},
\end{align}
where $N_\pm$ is the number of zero modes of $\tilde{H}^{\SIBC}$ with chirality $\pm 1$.
Hence, we get $N_+$ linearly independent eigenfunctions $\phi_n\in\mathcal{M}\otimes\ell^2(X)$ ($n=1,\ldots,N_+$) satisfying
\begin{align}
    \tilde{H}\modified{^{\SIBC}}{}\begin{pmatrix}
        \phi_n \\
        0
    \end{pmatrix}
    = 0,
    \quad
    \text{i.e.,}
    \quad
    H|_{\mathcal{M}}^{\SIBC}\phi_n
    = 0.
\end{align}
The zero modes $\phi_n$ of the restricted Hamiltonian $H|_{\mathcal{M}}^{\SIBC}$ are nothing but zero modes of $H^{\SIBC}$,
\begin{align}
    H^{\SIBC}\phi_n
    = H|_{\mathcal{M}}^{\SIBC}\phi_n
    = 0,
\end{align}
so $H^{\SIBC}$ hosts at least $N_+$ zero modes.
Therefore, we have the bulk-boundary correspondence
\begin{align}\label{eq:BBC}
    \#(\text{zero modes of }H^{\SIBC})
    \ge N_+
    \ge w^\mathcal{M}.
\end{align}
Apparently, Eq.~\eqref{eq:BBC} makes no sense if $w^\mathcal{M}<0$.
This is because we are considering SIBC systems on the positive side with respect to the boundary.
By taking another SIBC Hamiltonian $H^{-\SIBC}$ on $\mathbb{Z}_{\le 0}\times\mathbb{Z}^{d-1}$, we also have
\begin{align}
    \#(\text{zero modes of }H^{-\SIBC})
    \ge -w^\mathcal{M},
\end{align}
which makes sense for $w^{\mathcal{M}}<0$.

So far we assume the subspace property with $\dim\mathcal{M}=\dim\mathcal{M}'$.
As mentioned above, we can further divide it into two cases, $\mathcal{M}\neq\mathcal{M}'$ and $\mathcal{M}=\mathcal{M}'$.
If $\mathcal{M}\neq\mathcal{M}'$, the subspace property is unstable against energy shifts and thus protects only zero modes under the SIBC among all the boundary modes.
By contrast, if $\mathcal{M}=\mathcal{M}'$, we can shift the origin of energy respecting the subspace property, so $H^{(-)\SIBC}$ has at least $(-)w^{\mathcal{M}}$ modes at any energy $E\in\mathbb{C}$ in a neighborhood of zero, which are nothing but skin modes in one dimension \cite{Okuma2020}.
Ultimately, a nonzero subspace-protected topological invariant $w^\mathcal{M}$ implies two distinct boundary phenomena: the non-Hermitian skin effect for $\mathcal{M}=\mathcal{M}'$ and the presence of boundary zero modes for $\mathcal{M}\neq\mathcal{M}'$.

We expect the boundary phenomena in nontrivial phases to occur under the open boundary condition (OBC) as well as the SIBC.
We illustrate this expectation with several examples of $\mathcal{H}=\mathbb{C}^2$ and $\dim\mathcal{M}=\dim\mathcal{M}'=1$.
Without loss of generality, we take a specific basis where the basis vector of $\mathcal{M}$ is $\begin{pmatrix} 1\\0\end{pmatrix}$.
Then, let $\begin{pmatrix} a \\ b \end{pmatrix}$ be the basis vector of $\mathcal{M}'$.
Now the subspace property reads
$
    \begin{pmatrix}
        H_{11}(\boldsymbol{k}) \\
        H_{21}(\boldsymbol{k})
    \end{pmatrix}
    \propto \begin{pmatrix}
        a \\
        b
    \end{pmatrix}
$,
so there is a scalar function $h(\boldsymbol{k})$ of $\boldsymbol{k}$ such that 
\begin{align}
    H(\boldsymbol{k})
    = \begin{pmatrix}
        ah(\boldsymbol{k}) & H_{12}(\boldsymbol{k}) \\
        bh(\boldsymbol{k}) & H_{22}(\boldsymbol{k})
    \end{pmatrix}.
\end{align}
The restricted Hamiltonian is $H(\boldsymbol{k})|_{\mathcal{M}}=h(\boldsymbol{k})$.

\begin{figure}
    \centering
    \includegraphics[width=1.0\linewidth]{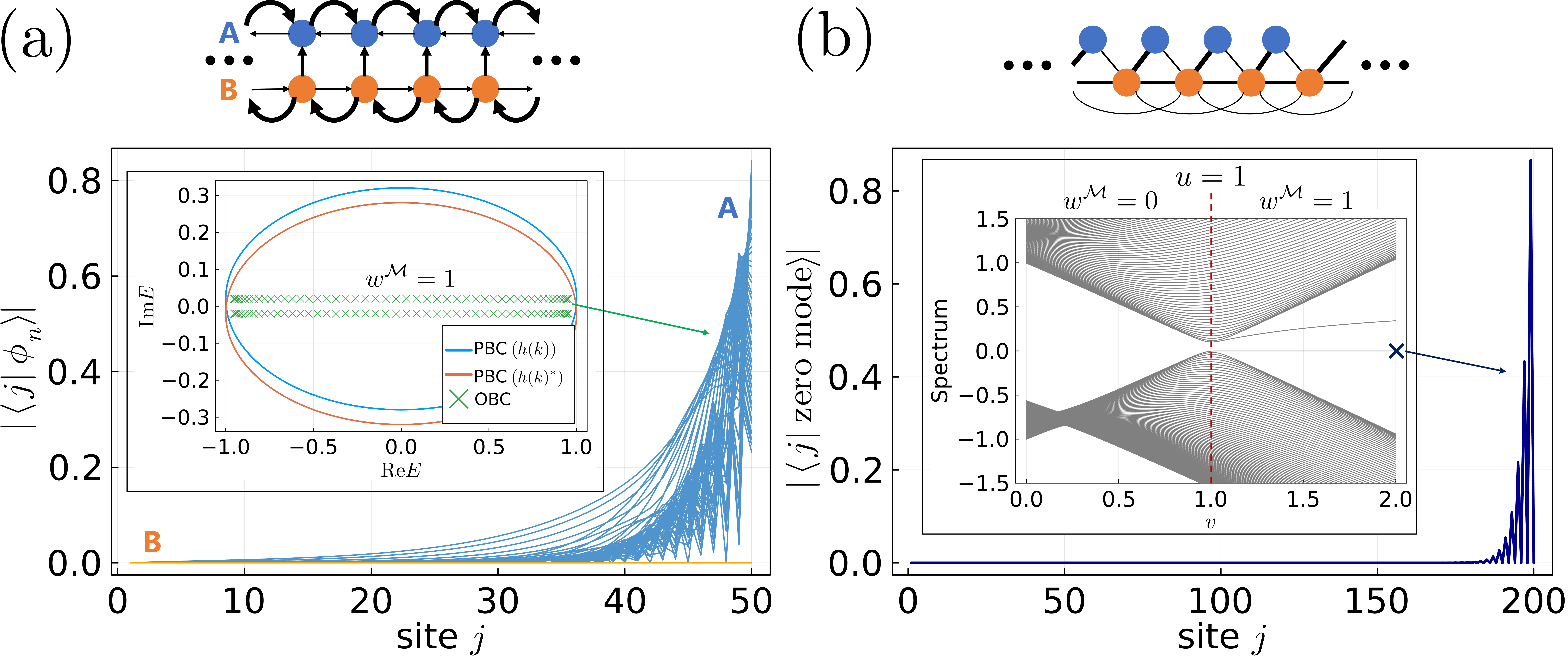}
    \caption{Two tight-binding models (top) and their bulk-boundary correspondence (bottom). (a) The one-way coupled model \eqref{eq:one-way_coupled_model} of two Hatano-Nelson chains ($t=1,g=0.3,V=0.02,r(k)=0.2$). The bottom figure depicts spatial distribution of all the eigenvectors under the OBC, which are localized at the right end in the sublattice A. The inset shows the PBC and OBC spectra.  (b) The extended SSH chain ($u=1,r(k)=0.6+0.5\cos k+0.2\sin 2k$). The bottom figure depicts the spatial distribution of a zero mode for $v=2$, which localized at the right end. The inset shows the OBC spectra for different values of $v$. A zero mode appears at $u=v$, reflecting the topological phase transition in the bulk from $w^{\mathcal{M}}=0$ to $w^{\mathcal{M}}=1$.}
    \label{fig:Simple_models}
\end{figure}

In the case of $\mathcal{M}=\mathcal{M}'$ with $a=1,b=0$, $H(\boldsymbol{k})$ becomes a triangular matrix as
\begin{align}
    H(\boldsymbol{k})
    = \begin{pmatrix}
        h(\boldsymbol{k}) & H_{12}(\boldsymbol{k}) \\
        0 & H_{22}(\boldsymbol{k})
    \end{pmatrix}.
\end{align}
As a simple model, we consider a one-dimensional non-Hermitian system given by
\begin{align}\label{eq:one-way_coupled_model}
    H(k)
    = \begin{pmatrix}
        h(k) & r(k) \\
        0 & h(k)^*
    \end{pmatrix},
    \quad
    h(k)
    = t\cos k + \ii g\sin k + \ii V,
\end{align}
with $t,g,V\in\mathbb{R}\modified{\setminus\{0\}}{^\times}$, $|V|<|g|$ and an arbitrary function $r(k)$.
Physically, this is a one-way coupled model of two Hatano-Nelson chains \cite{Hatano1996,Hatano1997} without disorder.
The spectrum of $H(k)$ is $\{h(k),h(k)^*\}$, which forms two ellipses.
While the prime point-gap topological invariant (the winding number) of $H(k)$ is zero
\begin{align}
    w
    = \frac{1}{2\pi\ii}\int_{\mathrm{BZ}}\Tr[H(k)^{-1}\dif H(k)]
    = 0,
\end{align}
the subspace-protected topological invariant is non-zero
\begin{align}
    w^{\mathcal{M}}
    = \frac{1}{2\pi\ii}\int_{\mathrm{BZ}}h(k)^{-1}\dif h(k)
    = \sgn\left(\frac{g}{t}\right)
    \neq 0.
\end{align}
Thereby, we expect the presence of the skin effect, exhibiting macroscopic number of boundary-localized eigenmodes under the OBC and distinct spectra under the periodic boundary condition (PBC) and OBC.
In Fig.~\ref{fig:Simple_models} (a), we provide the eigenvectors under the OBC and the PBC and OBC spectra of the model.
As shown in the figure, the skin modes appear under the OBC, despite $w=0$.  
The {\it zero winding skin modes} are consistent with the nonzero $w^{\mathcal{M}}$,
and thus are protected by the subspace property, beyond the conventional notion of symmetry.

As an example of $\mathcal{M}\neq\mathcal{M}'$, suppose that $a=0,b=1$, i.e., $\mathcal{M}'=\mathcal{M}^\perp$.
Then, $H(\boldsymbol{k})$ is of the form
\begin{align}
    H(\boldsymbol{k})
    = \begin{pmatrix}
        0 & H_{12}(\boldsymbol{k}) \\
        h(\boldsymbol{k}) & H_{22}(\boldsymbol{k})
    \end{pmatrix}.
\end{align}
In particular, we consider a one-dimensional Hermitian Hamiltonian given by
\begin{align}\label{eq:extended_SSH_chain}
    H(k)
    = \begin{pmatrix}
        0 & h(k)^* \\
        h(k) & r(k)
    \end{pmatrix},
    \quad
    h(k)
    = u+ve^{\ii k},
\end{align}
with $|u|\neq |v|$ to be point-gapped at zero energy.
\modified{A related model has been investigated in the context of sub-symmetry \cite{Wang2023}, but the idea of the subspace property facilitates understanding of bulk-boundary correspondence in the model.}{}
For $r=0$ this model is just the Su-Schrieffer-Heeger (SSH) chain \cite{Jackiw1976,Su1979,Su1980,Jackiw1981,Heeger1988} with the SLS, but for nonzero $r$ it no longer has any symmetry.
Still, we can define the subspace-protected topological invariant $w^{\mathcal{M}}$ for this model as
\begin{align}
    w^{\mathcal{M}}
    = \frac{1}{2\pi\ii}\int_{\mathrm{BZ}}h(k)^{-1}\dif h(k)
    = \begin{dcases*}
        1 & if $|u|<|v|$ \\
        0 & if $|u|>|v|$
    \end{dcases*},
\end{align}
independent of $r(k)$.
Corresponding to the topological phase transition with respect to $w^{\mathcal{M}}$, the boundary zero mode appears under the OBC, protected by the subspace property, as seen in Fig.~\ref{fig:Simple_models} (b).
In sharp contrast to the conventional SSH chain \modified{respecting the SLS}{}, the nontrivial phase supports an {\it unpaired zero mode} (a localized zero mode only at a single boundary), and \modified{the valence and conduction bands never cross}{the bulk gap never closes} even at the transition point. 

The bulk-boundary correspondence between $w^{\mathcal{M}}$ and boundary zero modes is always valid for Hermitian Hamiltonians with $\mathcal{M}\neq\mathcal{M}'$, because there is no skin effect.
However, if the system suffers from skin effects, say due to non-zero $w$ \cite{Okuma2020,Zhang2020}, $w^{\mathcal{M}}$ could fail to bound the number of boundary zero modes.
In this case, we should use a subspace-protected topological invariant defined on the generalized Brillouin zone (GBZ) to recover the bulk-boundary correspondence \cite{Yao2018}.

\textit{Interplay of subspace-protected topological phases and internal symmetries}.---
So far, we have considered subspace-protected topological phases in the absence of any symmetries.
Still, the subspace property together with additional symmetry brings a new topological phase.
We realize this by imposing additional symmetry to the restricted Hamiltonian $H(\boldsymbol{k})|_{\mathcal{M}}$.
Fixing the basis of $\mathcal{H}$ with a particular one, we obtain the matrix representation of $H(\boldsymbol{k})|_{\mathcal{M}}$, for which the Bernard-LeClair type of symmetry condition \cite{Bernard2002,Kawabata2019b} can be defined.
\modified{In such cases, subspace-protected topological invariants are defined as point-gap invariants for the restricted Hamiltonian, depending on symmetry classes and dimensionality.}{}

To be concrete, we take an example where $H(k)|_{\mathcal{M}}$ satisfies symmetry conditions of class BDI \cite{Altland1997} as follows:
\begin{gather}
    \label{eq:BDI_model}
    H(k)
    = \begin{pmatrix}
        0 & H_1(k)^\dagger \\
        H_1(k) & H_2(k)
    \end{pmatrix}, \\
    H_1(k)
    = \ii g\sin k + (h+t\cos k)\sigma_z + \ii r_x\sin k\sigma_x + r_y\sin k\sigma_y, \\
    H_2(k)
    = d_0+d'_0\sin k + \sum_{i=x,y,z}(d_i+d'_i\sin k)\sigma_i
\end{gather}
with $h,t,g,r_x,r_y\in\mathbb{R}$, $d_\mu,d'_\mu\in\modified{\mathbb{R}}{\mathbb{C}}$ ($\mu=0,x,y,z$).
The Hermitian Hamiltonian \eqref{eq:BDI_model} has no symmetry due to the presence of $H_2(k)$ but satisfies the subspace property with $\mathcal{M}'=\mathcal{M}^\perp=0\oplus 0\oplus\mathbb{C}\oplus\mathbb{C}$. 
The restricted Hamiltonian is $H(k)|_{\mathcal{M}}=H_1(k)$, which is non-Hermitian and has time reversal symmetry (TRS) $H(-k)|_{\mathcal{M}}^* = H(k)|_{\mathcal{M}}$, particle-hole symmetry (PHS) $\sigma_x H(-k)|_{\mathcal{M}}^{\mathsf{T}}\sigma_x = -H(k)|_{\mathcal{M}}$, and chiral symmetry (CS) $\sigma_x H(k)|_{\mathcal{M}}^\dagger\sigma_x = -H(k)|_{\mathcal{M}}$.
Whereas the PHS or CS for $H(k)|_{\mathcal{M}}$ makes the subspace-protected topological invariant $w^{\mathcal{M}}$ in Eq.~\eqref{eq:winding} trivial, the symmetries enable us to define another point-gap topological invariant $\nu\in\{0,1\}$ for $H(k)|_{\mathcal{M}}$ as \cite{Kawabata2019b}
\begin{align}
    (-1)^\nu
    \coloneqq \sgn\left[ \frac{\operatorname{Pf}(H(\pi)|_\mathcal{M}\sigma_x)}{\operatorname{Pf}(H(0)|_\mathcal{M}\sigma_x)} \right]
    = \operatorname{sgn}\left( \frac{h-t}{h+t} \right)
\end{align}
if $H(k)|_{\mathcal{M}}$ is point-gapped around $E=0$.

\begin{figure}
    \centering
    \includegraphics[width=1.0\linewidth]{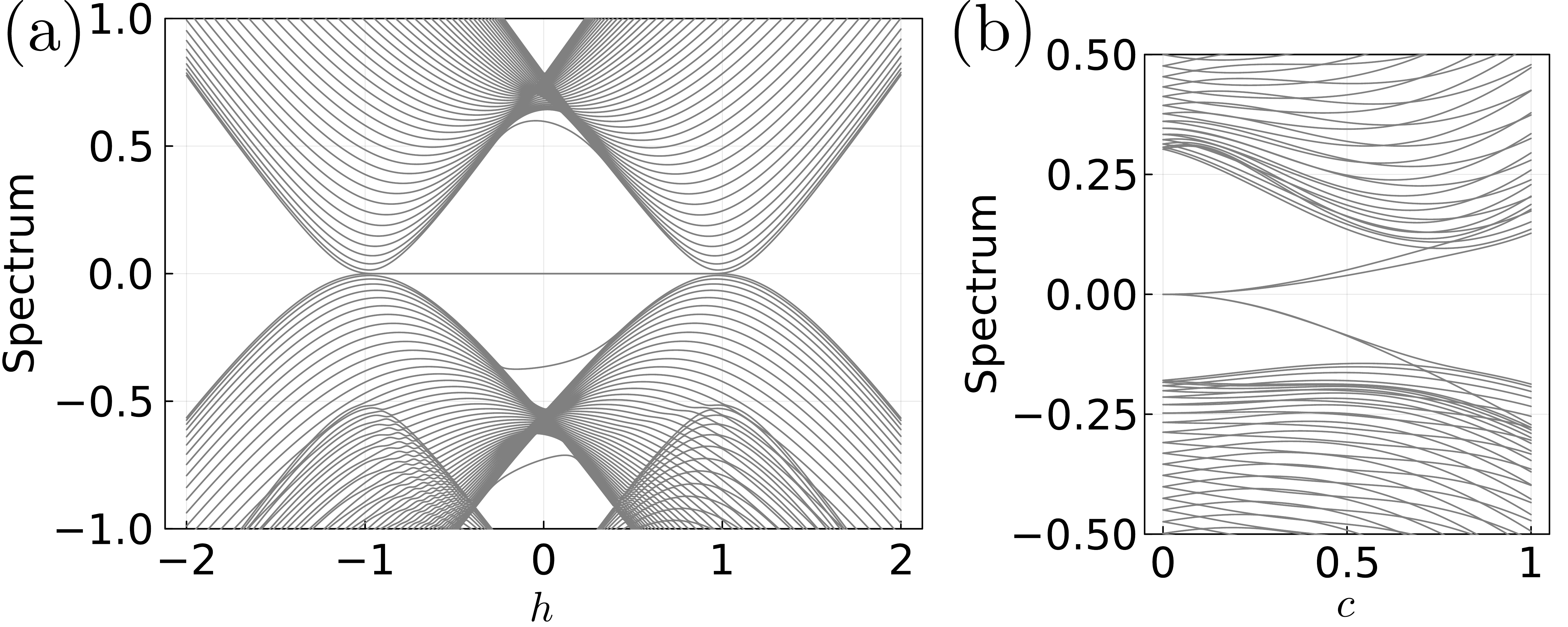}
    \caption{(a) Spectra of the Hermitian model \eqref{eq:BDI_model} for different values of $h$. The paremeters are chosen as $t=1,g=-0.9,r_x=-0.2,r_y=0.5,d_0=0.35,d_1=-0.086,d_2=0.22,d_3=0.83,d'_0=-0.12,d'_1=-0.44,d'_2=0.075,d'_3=-0.088$. In the nontrivial phases ($-1<h<1$) of $\nu=1$, there are two-fold degenerate zero modes. (b) Spectra of the stacked model $H_{\text{stack}}(k)$ in Eq.~\eqref{eq:stacked_model} for different values of $c$. The parameter $h$ is taken as $h=0.5$ and the others are chosen same as (a).}
    \label{fig:BDI}
\end{figure}

The bulk-boundary correspondence is still confirmed for the current case by constructing $\tilde{H}(k)$ in Eq.~\eqref{eq:doubled} again. 
As numerical evidence, we provide spectra of the model \eqref{eq:BDI_model} resolved by the parameter $h$ in Fig.~\ref{fig:BDI} (a). 
Topological phase transitions occur at $h=\pm t$, and the model hosts two-degenerate zero modes in the nontrivial phases ($|h|<|t|$) of $\nu=1$.

The presence of the $\mathbb{Z}_2$ invariant $\nu$ suggests that stacked models of $H(k)$ always belong to the trivial phase of $\nu=0$.
To check this, we stack two $H(k)$'s and add a coupling term $\Delta$ to construct the following model,
\begin{align}\label{eq:stacked_model}
    H_{\text{stack}}(k)
    = \begin{pmatrix}
        H(k) & \Delta^\dagger \\
        \Delta & H(k)
    \end{pmatrix},
    \quad
    \Delta
    = \begin{pmatrix}
        0 & cI_2 \\
        -cI_2 & 0
    \end{pmatrix},
    \quad
    c\in\mathbb{R},
\end{align}
with respecting the subspace property $H_{\text{stack}}(k)\mathcal{M}\oplus\mathcal{M}\subseteq\mathcal{M}'\oplus\mathcal{M}'$ and the symmetry conditions of class BDI for the restricted Hamiltonian $H_{\text{stack}}(k)|_{\mathcal{M}\oplus\mathcal{M}}=\begin{pmatrix} H_1(k) & cI_2 \\ -cI_2 & H_1(k) \end{pmatrix}$.
As shown in Fig.~\ref{fig:BDI} (b), the four-degenerate zero modes at $c=0$ are gapped out as the coupling strength $c$ increases.
Thus, the coexistence of the subspace property and symmetry condition for the restricted Hamiltonian $H(\boldsymbol{k})|_\mathcal{M}$ guarantees a $\mathbb{Z}_2$ topological phase, without any symmetry of the whole Hamiltonian $H(\boldsymbol{k})$.

\textit{Guideline for the search of physical platforms possessing the subspace property}.---
The subspace property may manifest itself in various physical systems.
In what follows, we give an overview of relevant physical platforms for cases $\mathcal{M}=\mathcal{M}'$ and $\mathcal{M}\perp\mathcal{M}'$.
\modified{A more detailed explanation is provided in the End Matter.}{}

One can realize the case $\mathcal{M}=\mathcal{M}'$ by coupling a system to another system unidirectionally as in Eq.~\eqref{eq:one-way_coupled_model} \cite{Zhong2020,Shi2022,Takami2025}.
Relevant experiments have already been carried out in an active mechanical lattice \cite{Li2025} and an electrical circuit \cite{Zhou2025}.
Such a system can host zero-winding skin modes with exceptional deficiency \cite{Li2025}.
In open quantum systems, furthermore, the self-energy of a fermionic Green function in the Keldysh formalism \cite{Larkin1975,Rammer2007,Altland2023} or the third-quantization form of a quadratic fermionic Lindbladian \cite{Prosen2008,Barthel2022,McDonald2023} is written by a triangular matrix, which has a subspace property with $\mathcal{M}=\mathcal{M}'$.
Thereby, we can define a subspace-protected topological invariant for the self-energy or Lindbladian, which depends only on the retarded and advanced Green functions or the effective Hamiltonian.

Moving on to the case $\mathcal{M}\perp\mathcal{M}'$, we have two approaches to find it in materials.
First, by partially breaking the sublattice symmetry of a bipartite system, the desired subspace property is achieved.
This approach shares the idea with the subsymmetry-protected topological phases \cite{Wang2023,Kang2024}.
An example is a bilayer graphene on a boron nitride substrate, where couplings to the substrate break the sublattice symmetry of a graphene in the lower layer \cite{McCann2013}.
Second approach is to hybridize a flat band with dispersive bands.
For instance, the effective Hamiltonian in equilibrium heavy fermion systems may \modified{be approximated by a matrix satisfying the}{possess an approximate} subspace property with $\mathcal{M}\perp\mathcal{M}'$ if both the self-energy and the $d$- or $f$-bands have little dispersion.

\textit{Discussion}.---
In summary, we have discovered subspace-protected topological phases and established the bulk-boundary correspondence.
Our work reveals that the new principle beyond symmetry may play an essential role in topological phases.

Conventional topological phases have a close relation to quantum anomalies \cite{Callan1985,Qi2008,Ryu2012,Hsieh2016,Witten2016,Witten2016b,Freed2021,Okuma2021,Kawabata2021,Nakai2025}, so we expect that some anomaly corresponds to subspace-protected topological phases from field-theoretical viewpoints.
To clarify a \textit{subspace anomaly}, we could refer to noninvertible symmetry \cite{Bhardwaj2018,Chang2019,Shao2024,Schafer-Nameki2024}, in which not a group but a category describes symmetry operations due to the absence of an inversion operation.
In fact, the subspace property can be regarded as a noninvertible generalization of some symmetries, such as unitary symmetry for $\mathcal{M}=\mathcal{M}'$ and sublattice symmetry for $\mathcal{M}'=\mathcal{M}^\perp$.
A subspace anomaly might yield novel phases of matter in strongly-correlated systems such as invertible phases or topologically ordered phases.

In the Letter, we have assumed in the subspace property \eqref{eq:subspace_property} that $\mathcal{M}$ and $\mathcal{M}'$ are complex subspaces.
Taking $\mathcal{M}$ and $\mathcal{M}'$ as real or quaternionic subspaces may enrich subspace-protected topological phases.
Furthermore, $K$-theoretical classification of the phases is an open problem.
\modified{
The concept of subspace-protected topological phases also may be extended to gapless systems to explain the stability of a type of semimetal.  
We will leave them for future research.}{}


\textit{Note added}.--- 
A part of this work was presented in \cite{Takami2025_APS}.
After completing this work, we became aware of a recent work~\cite{Ma2025}. Although Ref.~\cite{Ma2025} introduces a similar terminology "\textit{subspace symmetry}", it is an emergent symmetry of a Hamiltonian, not a property of a subspace. Thus, the concept is entirely different from ours.

\medskip
\begingroup
\renewcommand{\addcontentsline}[3]{}
\begin{acknowledgments}
We thank Guancong Ma for fruitful discussions.
K.S. thanks Frank Schindler, Tijan Prijon, Shu Hamanaka, and Tsuneya Yoshida for valuable discussions.
K.S. and M.S. thank Jan Wiersig for informing us of related works.
We are supported by JST CREST Grant No.~JPMJCR19T2. 
K.S. is supported by JSPS KAKENHI Grant No.~JP25KJ1632. 
D.N. is supported by JSPS KAKENHI Grant No.~JP24K22857.
M.S. is supported by JSPS KAKENHI Grant Nos.~JP24K00569 and JP25H01250. 
\end{acknowledgments}
\endgroup

\let\oldaddcontentsline\addcontentsline
\renewcommand{\addcontentsline}[3]{}
\bibliography{bib.bib}

@article{Zhou2025,
  title = {Observation of anomalous non-Hermitian skin effect in electric circuits},
  author = {Zhou, Dongmei and Nie, Lang and Hu, Ruixia and Wang, Xiaoyun and Wu, Jien and He, Zhaojian and Deng, Ke},
  journal = {Phys. Rev. B},
  volume = {111},
  issue = {22},
  pages = {224104},
  numpages = {10},
  year = {2025},
  month = {Jun},
  publisher = {American Physical Society},
  doi = {10.1103/PhysRevB.111.224104},
  url = {https://link.aps.org/doi/10.1103/PhysRevB.111.224104}
}

@Article{Guo2023,
author={Guo, Gang-Feng
and Bao, Xi-Xi
and Zhu, Han-Jie
and Zhao, Xiao-Ming
and Zhuang, Lin
and Tan, Lei
and Liu, Wu-Ming},
title={Anomalous non-Hermitian skin effect: topological inequivalence of skin modes versus point gap},
journal={Communications Physics},
year={2023},
month={Dec},
day={19},
volume={6},
number={1},
pages={363},
issn={2399-3650},
doi={10.1038/s42005-023-01487-4},
url={https://doi.org/10.1038/s42005-023-01487-4},
  archivePrefix = {arXiv},
  eprint = {2304.06926},
  primaryClass = {cond-mat},
}

@misc{Li2025,
      title={Exceptional deficiency of non-Hermitian systems: high-dimensional coalescence and dynamics}, 
      author={Zhen Li and Xulong Wang and Rundong Cai and Kenji Shimomura and Zhesen Yang and Masatoshi Sato and Guancong Ma},
      year={2025},
      eprint={2504.12238},
      archivePrefix={arXiv},
      primaryClass={quant-ph},
      url={https://arxiv.org/abs/2504.12238}, 
}

@Article{Wang2023,
author={Wang, Ziteng
and Wang, Xiangdong
and Hu, Zhichan
and Bongiovanni, Domenico
and Juki{\'{c}}, Dario
and Tang, Liqin
and Song, Daohong
and Morandotti, Roberto
and Chen, Zhigang
and Buljan, Hrvoje},
title={Sub-symmetry-protected topological states},
journal={Nature Physics},
year={2023},
month={Jul},
day={01},
volume={19},
number={7},
pages={992-998},
issn={1745-2481},
doi={10.1038/s41567-023-02011-9},
url={https://doi.org/10.1038/s41567-023-02011-9},
  archivePrefix = {arXiv},
  eprint = {2205.07285},
  primaryClass = {cond-mat},
}

@article{Verma2024,
  title = {Bulk-boundary correspondence in extended trimer Su-Schrieffer-Heeger model},
  author = {Verma, Sonu and Ghosh, Tarun Kanti},
  journal = {Phys. Rev. B},
  volume = {110},
  issue = {12},
  pages = {125424},
  numpages = {15},
  year = {2024},
  month = {9},
  publisher = {American Physical Society},
  doi = {10.1103/PhysRevB.110.125424},
  url = {https://link.aps.org/doi/10.1103/PhysRevB.110.125424},
  archivePrefix = {arXiv},
  eprint = {2401.11695},
  primaryClass = {cond-mat},
}

@article{Verma2024b,
  title = {Non-Bloch band theory of subsymmetry-protected topological phases},
  author = {Verma, Sonu and Park, Moon Jip},
  journal = {Phys. Rev. B},
  volume = {110},
  issue = {3},
  pages = {035424},
  numpages = {14},
  year = {2024},
  month = {Jul},
  publisher = {American Physical Society},
  doi = {10.1103/PhysRevB.110.035424},
  url = {https://link.aps.org/doi/10.1103/PhysRevB.110.035424},
  archivePrefix = {arXiv},
  eprint = {2405.06240},
  primaryClass = {cond-mat},
}

@article{Kang2024,
  title = {Subsymmetry protected topology in topological insulators and superconductors},
  author = {Kang, Myungjun and Lee, Mingyu and Cheon, Sangmo},
  journal = {Phys. Rev. Res.},
  volume = {6},
  issue = {3},
  pages = {033323},
  numpages = {9},
  year = {2024},
  month = {Sep},
  publisher = {American Physical Society},
  doi = {10.1103/PhysRevResearch.6.033323},
  url = {https://link.aps.org/doi/10.1103/PhysRevResearch.6.033323},
  archivePrefix = {arXiv},
  eprint = {2406.01089},
  primaryClass = {cond-mat},
}

@article{Liu2023,
  title = {Analytic solution of the $n$-dimensional Su-Schrieffer-Heeger model},
  author = {Liu, Feng},
  journal = {Phys. Rev. B},
  volume = {108},
  issue = {24},
  pages = {245140},
  numpages = {10},
  year = {2023},
  month = {Dec},
  publisher = {American Physical Society},
  doi = {10.1103/PhysRevB.108.245140},
  url = {https://link.aps.org/doi/10.1103/PhysRevB.108.245140},
  archivePrefix = {arXiv},
  eprint = {2304.11933},
  primaryClass = {cond-mat},
}

@article{Jangjan2022,
  title = {Topological properties of subsystem-symmetry-protected edge states in an extended quasi-one-dimensional dimerized lattice},
  author = {Jangjan, Milad and Hosseini, Mir Vahid},
  journal = {Phys. Rev. B},
  volume = {106},
  issue = {20},
  pages = {205111},
  numpages = {9},
  year = {2022},
  month = {Nov},
  publisher = {American Physical Society},
  doi = {10.1103/PhysRevB.106.205111},
  url = {https://link.aps.org/doi/10.1103/PhysRevB.106.205111},
  archivePrefix = {arXiv},
  eprint = {2203.13160},
  primaryClass = {cond-mat},
}

@article{Sato2011,
  title = {Topology of Andreev bound states with flat dispersion},
  author = {Sato, Masatoshi and Tanaka, Yukio and Yada, Keiji and Yokoyama, Takehito},
  journal = {Phys. Rev. B},
  volume = {83},
  issue = {22},
  pages = {224511},
  numpages = {22},
  year = {2011},
  month = {Jun},
  publisher = {American Physical Society},
  doi = {10.1103/PhysRevB.83.224511},
  url = {https://link.aps.org/doi/10.1103/PhysRevB.83.224511},
  archivePrefix = {arXiv},
  eprint = {1102.1322},
  primaryClass = {cond-mat},
}

@book{Prodan2016,
  title={Bulk and Boundary Invariants for Complex Topological Insulators: From K-Theory to Physics},
  author={Prodan, E. and Schulz-Baldes, H.},
  isbn={9783319293516},
  series={Mathematical Physics Studies},
  url={https://books.google.co.jp/books?id=_-OKCwAAQBAJ},
  year={2016},
  publisher={Springer International Publishing},
  archivePrefix = {arXiv},
  eprint = {1510.08744},
  primaryClass = {math-ph},
}

@article{Ma2025,
  title = {Subspace symmetry for Bloch electrons},
  author = {Ma, Hai-Yang and Zhang, Shihao and Zhao, Yuanyuan and Jiang, Zhicheng and Xu, Hu and Jia, Jin-Feng},
  journal = {Phys. Rev. B},
  volume = {112},
  issue = {9},
  pages = {094105},
  numpages = {5},
  year = {2025},
  month = {9},
  publisher = {American Physical Society},
  doi = {10.1103/vypg-my88},
  url = {https://link.aps.org/doi/10.1103/vypg-my88}
}

@article{Wen2017,
  title = {Colloquium: Zoo of quantum-topological phases of matter},
  author = {Wen, Xiao-Gang},
  journal = {Rev. Mod. Phys.},
  volume = {89},
  issue = {4},
  pages = {041004},
  numpages = {17},
  year = {2017},
  month = {12},
  publisher = {American Physical Society},
  doi = {10.1103/RevModPhys.89.041004},
  url = {https://link.aps.org/doi/10.1103/RevModPhys.89.041004},
  archivePrefix = {arXiv},
  eprint = {1610.03911},
  primaryClass = {cond-mat},
}

@article{Hatsugai1993,
  title = {Edge states in the integer quantum Hall effect and the Riemann surface of the Bloch function},
  author = {Hatsugai, Yasuhiro},
  journal = {Phys. Rev. B},
  volume = {48},
  issue = {16},
  pages = {11851--11862},
  numpages = {0},
  year = {1993},
  month = {10},
  publisher = {American Physical Society},
  doi = {10.1103/PhysRevB.48.11851},
  url = {https://link.aps.org/doi/10.1103/PhysRevB.48.11851}
}

@article{Hatsugai1993b,
  title = {Chern number and edge states in the integer quantum Hall effect},
  author = {Hatsugai, Yasuhiro},
  journal = {Phys. Rev. Lett.},
  volume = {71},
  issue = {22},
  pages = {3697--3700},
  numpages = {0},
  year = {1993},
  month = {11},
  publisher = {American Physical Society},
  doi = {10.1103/PhysRevLett.71.3697},
  url = {https://link.aps.org/doi/10.1103/PhysRevLett.71.3697}
}

@Article{Elbau2002,
author={Elbau, P.
and Graf, G. M.},
title={Equality of Bulk and Edge Hall Conductance Revisited},
journal={Communications in Mathematical Physics},
year={2002},
month={9},
day={01},
volume={229},
number={3},
pages={415--432},
issn={1432-0916},
doi={10.1007/s00220-002-0698-z},
url={https://doi.org/10.1007/s00220-002-0698-z},
  archivePrefix = {arXiv},
  eprint = {math-ph/0203019},
}

@article{Mong2011,
  title = {Edge states and the bulk-boundary correspondence in Dirac Hamiltonians},
  author = {Mong, Roger S. K. and Shivamoggi, Vasudha},
  journal = {Phys. Rev. B},
  volume = {83},
  issue = {12},
  pages = {125109},
  numpages = {15},
  year = {2011},
  month = {3},
  publisher = {American Physical Society},
  doi = {10.1103/PhysRevB.83.125109},
  url = {https://link.aps.org/doi/10.1103/PhysRevB.83.125109},
  archivePrefix = {arXiv},
  eprint = {1010.2778},
  primaryClass = {cond-mat},
}

@article{Essin2011,
  title = {Bulk-boundary correspondence of topological insulators from their respective Green's functions},
  author = {Essin, Andrew M. and Gurarie, Victor},
  journal = {Phys. Rev. B},
  volume = {84},
  issue = {12},
  pages = {125132},
  numpages = {10},
  year = {2011},
  month = {9},
  publisher = {American Physical Society},
  doi = {10.1103/PhysRevB.84.125132},
  url = {https://link.aps.org/doi/10.1103/PhysRevB.84.125132},
  archivePrefix = {arXiv},
  eprint = {1104.1602},
  primaryClass = {cond-mat},
}

@Article{Graf2013,
author={Graf, Gian Michele
and Porta, Marcello},
title={Bulk-Edge Correspondence for Two-Dimensional Topological Insulators},
journal={Communications in Mathematical Physics},
year={2013},
month={12},
day={01},
volume={324},
number={3},
pages={851-895},
issn={1432-0916},
doi={10.1007/s00220-013-1819-6},
url={https://doi.org/10.1007/s00220-013-1819-6},
  archivePrefix = {arXiv},
  eprint = {1207.5989},
  primaryClass = {cond-mat},
}

@article{Witten2016,
  title={Three lectures on topological phases of matter},
  author={Witten, Edward},
  journal={La Rivista del Nuovo Cimento},
  volume={39},
  number={7},
  pages={313--370},
  year={2016},
  publisher={Springer},
  archivePrefix = {arXiv},
  eprint = {1510.07698},
  primaryClass = {cond-mat},
}

@article{Ryu2002,
  title = {Topological Origin of Zero-Energy Edge States in Particle-Hole Symmetric Systems},
  author = {Ryu, Shinsei and Hatsugai, Yasuhiro},
  journal = {Phys. Rev. Lett.},
  volume = {89},
  issue = {7},
  pages = {077002},
  numpages = {4},
  year = {2002},
  month = {7},
  publisher = {American Physical Society},
  doi = {10.1103/PhysRevLett.89.077002},
  url = {https://link.aps.org/doi/10.1103/PhysRevLett.89.077002},
  archivePrefix = {arXiv},
  eprint = {cond-mat/0112197},
}

@article{Bernevig2005,
  title = {Intrinsic Spin Hall Effect in the Two-Dimensional Hole Gas},
  author = {Bernevig, B. Andrei and Zhang, Shou-Cheng},
  journal = {Phys. Rev. Lett.},
  volume = {95},
  issue = {1},
  pages = {016801},
  numpages = {4},
  year = {2005},
  month = {6},
  publisher = {American Physical Society},
  doi = {10.1103/PhysRevLett.95.016801},
  url = {https://link.aps.org/doi/10.1103/PhysRevLett.95.016801},
  archivePrefix = {arXiv},
  eprint = {cond-mat/0411457},
}

@article{Kane2005,
  title = {${Z}_{2}$ Topological Order and the Quantum Spin Hall Effect},
  author = {Kane, C. L. and Mele, E. J.},
  journal = {Phys. Rev. Lett.},
  volume = {95},
  issue = {14},
  pages = {146802},
  numpages = {4},
  year = {2005},
  month = {9},
  publisher = {American Physical Society},
  doi = {10.1103/PhysRevLett.95.146802},
  url = {https://link.aps.org/doi/10.1103/PhysRevLett.95.146802},
  archivePrefix = {arXiv},
  eprint = {cond-mat/0506581},
}

@article{Kane2005b,
  title = {Quantum Spin Hall Effect in Graphene},
  author = {Kane, C. L. and Mele, E. J.},
  journal = {Phys. Rev. Lett.},
  volume = {95},
  issue = {22},
  pages = {226801},
  numpages = {4},
  year = {2005},
  month = {11},
  publisher = {American Physical Society},
  doi = {10.1103/PhysRevLett.95.226801},
  url = {https://link.aps.org/doi/10.1103/PhysRevLett.95.226801},
  archivePrefix = {arXiv},
  eprint = {cond-mat/0411737},
}

@article{Bernevig2006,
  title = {Quantum Spin Hall Effect},
  author = {Bernevig, B. Andrei and Zhang, Shou-Cheng},
  journal = {Phys. Rev. Lett.},
  volume = {96},
  issue = {10},
  pages = {106802},
  numpages = {4},
  year = {2006},
  month = {3},
  publisher = {American Physical Society},
  doi = {10.1103/PhysRevLett.96.106802},
  url = {https://link.aps.org/doi/10.1103/PhysRevLett.96.106802},
  archivePrefix = {arXiv},
  eprint = {cond-mat/0504147},
}

@article{Fu2007,
  title = {Topological Insulators in Three Dimensions},
  author = {Fu, Liang and Kane, C. L. and Mele, E. J.},
  journal = {Phys. Rev. Lett.},
  volume = {98},
  issue = {10},
  pages = {106803},
  numpages = {4},
  year = {2007},
  month = {3},
  publisher = {American Physical Society},
  doi = {10.1103/PhysRevLett.98.106803},
  url = {https://link.aps.org/doi/10.1103/PhysRevLett.98.106803},
  archivePrefix = {arXiv},
  eprint = {cond-mat/0607699},
}

@article{Moore2007,
  title = {Topological invariants of time-reversal-invariant band structures},
  author = {Moore, J. E. and Balents, L.},
  journal = {Phys. Rev. B},
  volume = {75},
  issue = {12},
  pages = {121306},
  numpages = {4},
  year = {2007},
  month = {3},
  publisher = {American Physical Society},
  doi = {10.1103/PhysRevB.75.121306},
  url = {https://link.aps.org/doi/10.1103/PhysRevB.75.121306},
  archivePrefix = {arXiv},
  eprint = {cond-mat/0607314},
}

@article{Teo2008,
  title = {Surface states and topological invariants in three-dimensional topological insulators: Application to ${\text{Bi}}_{1\ensuremath{-}x}{\text{Sb}}_{x}$},
  author = {Teo, Jeffrey C. Y. and Fu, Liang and Kane, C. L.},
  journal = {Phys. Rev. B},
  volume = {78},
  issue = {4},
  pages = {045426},
  numpages = {15},
  year = {2008},
  month = {7},
  publisher = {American Physical Society},
  doi = {10.1103/PhysRevB.78.045426},
  url = {https://link.aps.org/doi/10.1103/PhysRevB.78.045426}
}

@article{Read2000,
  title = {Paired states of fermions in two dimensions with breaking of parity and time-reversal symmetries and the fractional quantum Hall effect},
  author = {Read, N. and Green, Dmitry},
  journal = {Phys. Rev. B},
  volume = {61},
  issue = {15},
  pages = {10267--10297},
  numpages = {0},
  year = {2000},
  month = {4},
  publisher = {American Physical Society},
  doi = {10.1103/PhysRevB.61.10267},
  url = {https://link.aps.org/doi/10.1103/PhysRevB.61.10267},
  archivePrefix = {arXiv},
  eprint = {cond-mat/9906453},
}

@article{Ivanov2001,
  title = {Non-Abelian Statistics of Half-Quantum Vortices in $\mathit{p}$-Wave Superconductors},
  author = {Ivanov, D. A.},
  journal = {Phys. Rev. Lett.},
  volume = {86},
  issue = {2},
  pages = {268--271},
  numpages = {0},
  year = {2001},
  month = {Jan},
  publisher = {American Physical Society},
  doi = {10.1103/PhysRevLett.86.268},
  url = {https://link.aps.org/doi/10.1103/PhysRevLett.86.268},
  archivePrefix = {arXiv},
  eprint = {cond-mat/0005069},
}

@article{Kitaev2001,
  title = {Unpaired Majorana fermions in quantum wires},
  author = {Kitaev, A. Yu.},
  journal = {Phys. Usp.},
  volume = {44},
  issue = {10 suppl.},
  pages = {s131--s136},
  year = {2001},
  doi = {10.1070/1063-7869/44/10S/S29},
  url = {https://doi.org/10.1070/1063-7869/44/10S/S29},
  archivePrefix = {arXiv},
  eprint = {cond-mat/0010440},
}

@article{Sato2003,
title = {Non-Abelian statistics of axion strings},
journal = {Physics Letters B},
volume = {575},
number = {1},
pages = {126-130},
year = {2003},
issn = {0370-2693},
doi = {https://doi.org/10.1016/j.physletb.2003.09.047},
url = {https://www.sciencedirect.com/science/article/pii/S037026930301428X},
author = {Sato, Masatoshi},
  archivePrefix = {arXiv},
  eprint = {hep-th/0307005},
}

@article{Fu2008,
  title = {Superconducting Proximity Effect and Majorana Fermions at the Surface of a Topological Insulator},
  author = {Fu, Liang and Kane, C. L.},
  journal = {Phys. Rev. Lett.},
  volume = {100},
  issue = {9},
  pages = {096407},
  numpages = {4},
  year = {2008},
  month = {3},
  publisher = {American Physical Society},
  doi = {10.1103/PhysRevLett.100.096407},
  url = {https://link.aps.org/doi/10.1103/PhysRevLett.100.096407},
  archivePrefix = {arXiv},
  eprint = {0707.1692},
  primaryClass = {cond-mat},
}

@book{Volovik2009,
    author = {Volovik, Grigory E.},
    title = {The Universe in a Helium Droplet},
    publisher = {Oxford University Press},
    year = {2009},
    month = {2},
    isbn = {9780199564842},
    doi = {10.1093/acprof:oso/9780199564842.001.0001},
    url = {https://doi.org/10.1093/acprof:oso/9780199564842.001.0001},
}

@article{Qi2009,
  title = {Time-Reversal-Invariant Topological Superconductors and Superfluids in Two and Three Dimensions},
  author = {Qi, Xiao-Liang and Hughes, Taylor L. and Raghu, S. and Zhang, Shou-Cheng},
  journal = {Phys. Rev. Lett.},
  volume = {102},
  issue = {18},
  pages = {187001},
  numpages = {4},
  year = {2009},
  month = {5},
  publisher = {American Physical Society},
  doi = {10.1103/PhysRevLett.102.187001},
  url = {https://link.aps.org/doi/10.1103/PhysRevLett.102.187001},
  archivePrefix = {arXiv},
  eprint = {0803.3614},
  primaryClass = {cond-mat},
}

@article{Sato2009,
  title = {Non-Abelian Topological Order in $s$-Wave Superfluids of Ultracold Fermionic Atoms},
  author = {Sato, Masatoshi and Takahashi, Yoshiro and Fujimoto, Satoshi},
  journal = {Phys. Rev. Lett.},
  volume = {103},
  issue = {2},
  pages = {020401},
  numpages = {4},
  year = {2009},
  month = {7},
  publisher = {American Physical Society},
  doi = {10.1103/PhysRevLett.103.020401},
  url = {https://link.aps.org/doi/10.1103/PhysRevLett.103.020401},
  archivePrefix = {arXiv},
  eprint = {0901.4693},
  primaryClass = {cond-mat},
}

@article{Sau2010,
  title = {Generic New Platform for Topological Quantum Computation Using Semiconductor Heterostructures},
  author = {Sau, Jay D. and Lutchyn, Roman M. and Tewari, Sumanta and Das Sarma, S.},
  journal = {Phys. Rev. Lett.},
  volume = {104},
  issue = {4},
  pages = {040502},
  numpages = {4},
  year = {2010},
  month = {1},
  publisher = {American Physical Society},
  doi = {10.1103/PhysRevLett.104.040502},
  url = {https://link.aps.org/doi/10.1103/PhysRevLett.104.040502},
  archivePrefix = {arXiv},
  eprint = {0907.2239},
  primaryClass = {cond-mat},
}

@article{Lutchyn2010,
  title = {Majorana Fermions and a Topological Phase Transition in Semiconductor-Superconductor Heterostructures},
  author = {Lutchyn, Roman M. and Sau, Jay D. and Das Sarma, S.},
  journal = {Phys. Rev. Lett.},
  volume = {105},
  issue = {7},
  pages = {077001},
  numpages = {4},
  year = {2010},
  month = {8},
  publisher = {American Physical Society},
  doi = {10.1103/PhysRevLett.105.077001},
  url = {https://link.aps.org/doi/10.1103/PhysRevLett.105.077001},
  archivePrefix = {arXiv},
  eprint = {1002.4033},
  primaryClass = {cond-mat},
}

@article{Alicea2010,
  title = {Majorana fermions in a tunable semiconductor device},
  author = {Alicea, Jason},
  journal = {Phys. Rev. B},
  volume = {81},
  issue = {12},
  pages = {125318},
  numpages = {10},
  year = {2010},
  month = {3},
  publisher = {American Physical Society},
  doi = {10.1103/PhysRevB.81.125318},
  url = {https://link.aps.org/doi/10.1103/PhysRevB.81.125318},
  archivePrefix = {arXiv},
  eprint = {0912.2115},
  primaryClass = {cond-mat},
}

@article{Oreg2010,
  title = {Helical Liquids and Majorana Bound States in Quantum Wires},
  author = {Oreg, Yuval and Refael, Gil and von Oppen, Felix},
  journal = {Phys. Rev. Lett.},
  volume = {105},
  issue = {17},
  pages = {177002},
  numpages = {4},
  year = {2010},
  month = {10},
  publisher = {American Physical Society},
  doi = {10.1103/PhysRevLett.105.177002},
  url = {https://link.aps.org/doi/10.1103/PhysRevLett.105.177002},
  archivePrefix = {arXiv},
  eprint = {1003.1145},
  primaryClass = {cond-mat},
}

@Article{Moore2010,
author={Moore, Joel E.},
title={The birth of topological insulators},
journal={Nature},
year={2010},
month={3},
day={01},
volume={464},
number={7286},
pages={194--198},
issn={1476-4687},
doi={10.1038/nature08916},
url={https://doi.org/10.1038/nature08916}
}

@article{Hasan2010,
  title = {Colloquium: Topological insulators},
  author = {Hasan, M. Z. and Kane, C. L.},
  journal = {Rev. Mod. Phys.},
  volume = {82},
  issue = {4},
  pages = {3045--3067},
  numpages = {0},
  year = {2010},
  month = {11},
  publisher = {American Physical Society},
  doi = {10.1103/RevModPhys.82.3045},
  url = {https://link.aps.org/doi/10.1103/RevModPhys.82.3045},
  archivePrefix = {arXiv},
  eprint = {1002.3895},
  primaryClass = {cond-mat},
}

@article{Qi2011,
  title = {Topological insulators and superconductors},
  author = {Qi, Xiao-Liang and Zhang, Shou-Cheng},
  journal = {Rev. Mod. Phys.},
  volume = {83},
  issue = {4},
  pages = {1057--1110},
  numpages = {0},
  year = {2011},
  month = {10},
  publisher = {American Physical Society},
  doi = {10.1103/RevModPhys.83.1057},
  url = {https://link.aps.org/doi/10.1103/RevModPhys.83.1057},
  archivePrefix = {arXiv},
  eprint = {1008.2026},
  primaryClass = {cond-mat},
}

@article{Chiu2016,
  title = {Classification of topological quantum matter with symmetries},
  author = {Chiu, Ching-Kai and Teo, Jeffrey C. Y. and Schnyder, Andreas P. and Ryu, Shinsei},
  journal = {Rev. Mod. Phys.},
  volume = {88},
  issue = {3},
  pages = {035005},
  numpages = {63},
  year = {2016},
  month = {8},
  publisher = {American Physical Society},
  doi = {10.1103/RevModPhys.88.035005},
  url = {https://link.aps.org/doi/10.1103/RevModPhys.88.035005},
  archivePrefix = {arXiv},
  eprint = {1505.03535},
  primaryClass = {cond-mat},
}

@article{Sato2017,
doi = {10.1088/1361-6633/aa6ac7},
url = {https://dx.doi.org/10.1088/1361-6633/aa6ac7},
year = {2017},
month = {5},
publisher = {IOP Publishing},
volume = {80},
number = {7},
pages = {076501},
author = {Sato, Masatoshi and Ando, Yoichi},
title = {Topological superconductors: a review},
journal = {Reports on Progress in Physics},
  archivePrefix = {arXiv},
  eprint = {1608.03395},
  primaryClass = {cond-mat},
}

@article{Schnyder2008,
  title = {Classification of topological insulators and superconductors in three spatial dimensions},
  author = {Schnyder, Andreas P. and Ryu, Shinsei and Furusaki, Akira and Ludwig, Andreas W. W.},
  journal = {Phys. Rev. B},
  volume = {78},
  issue = {19},
  pages = {195125},
  numpages = {22},
  year = {2008},
  month = {11},
  publisher = {American Physical Society},
  doi = {10.1103/PhysRevB.78.195125},
  url = {https://link.aps.org/doi/10.1103/PhysRevB.78.195125},
  archivePrefix = {arXiv},
  eprint = {0803.2786},
  primaryClass = {cond-mat},
}

@article{Schnyder2009,
    author = {Schnyder, Andreas P. and Ryu, Shinsei and Furusaki, Akira and Ludwig, Andreas W. W.},
    title = {Classification of Topological Insulators and Superconductors},
    journal = {AIP Conference Proceedings},
    volume = {1134},
    number = {1},
    pages = {10--21},
    year = {2009},
    month = {5},
    issn = {0094-243X},
    doi = {10.1063/1.3149481},
    url = {https://doi.org/10.1063/1.3149481},
  archivePrefix = {arXiv},
  eprint = {0905.2029},
  primaryClass = {cond-mat},
}

@article{Kitaev2009,
    author = {Kitaev, Alexei},
    title = {Periodic table for topological insulators and superconductors},
    journal = {AIP Conference Proceedings},
    volume = {1134},
    number = {1},
    pages = {22--30},
    year = {2009},
    month = {5},
    issn = {0094-243X},
    doi = {10.1063/1.3149495},
    url = {https://doi.org/10.1063/1.3149495},
  archivePrefix = {arXiv},
  eprint = {0901.2686},
  primaryClass = {cond-mat},
}

@article{Ryu2010,
doi = {10.1088/1367-2630/12/6/065010},
url = {https://dx.doi.org/10.1088/1367-2630/12/6/065010},
year = {2010},
month = {6},
publisher = {},
volume = {12},
number = {6},
pages = {065010},
author = {Ryu, Shinsei and Schnyder, Andreas P and Furusaki, Akira and Ludwig, Andreas W W},
title = {Topological insulators and superconductors: tenfold way and dimensional hierarchy},
journal = {New Journal of Physics},
  archivePrefix = {arXiv},
  eprint = {0912.2157},
  primaryClass = {cond-mat},
}

@article{Teo2010,
  title = {Topological defects and gapless modes in insulators and superconductors},
  author = {Teo, Jeffrey C. Y. and Kane, C. L.},
  journal = {Phys. Rev. B},
  volume = {82},
  issue = {11},
  pages = {115120},
  numpages = {26},
  year = {2010},
  month = {9},
  publisher = {American Physical Society},
  doi = {10.1103/PhysRevB.82.115120},
  url = {https://link.aps.org/doi/10.1103/PhysRevB.82.115120},
  archivePrefix = {arXiv},
  eprint = {1006.0690},
  primaryClass = {cond-mat},
}

@Article{Freed2013,
author={Freed, Daniel S.
and Moore, Gregory W.},
title={Twisted Equivariant Matter},
journal={Annales Henri Poincar{\'e}},
year={2013},
month={12},
day={01},
volume={14},
number={8},
pages={1927--2023},
issn={1424-0661},
doi={10.1007/s00023-013-0236-x},
url={https://doi.org/10.1007/s00023-013-0236-x},
  archivePrefix = {arXiv},
  eprint = {1208.5055},
  primaryClass = {cond-mat},
}

@article{Fu2007b,
  title = {Topological insulators with inversion symmetry},
  author = {Fu, Liang and Kane, C. L.},
  journal = {Phys. Rev. B},
  volume = {76},
  issue = {4},
  pages = {045302},
  numpages = {17},
  year = {2007},
  month = {7},
  publisher = {American Physical Society},
  doi = {10.1103/PhysRevB.76.045302},
  url = {https://link.aps.org/doi/10.1103/PhysRevB.76.045302},
  archivePrefix = {arXiv},
  eprint = {cond-mat/0611341},
}

@article{Roy2009,
  title = {Topological phases and the quantum spin Hall effect in three dimensions},
  author = {Roy, Rahul},
  journal = {Phys. Rev. B},
  volume = {79},
  issue = {19},
  pages = {195322},
  numpages = {5},
  year = {2009},
  month = {5},
  publisher = {American Physical Society},
  doi = {10.1103/PhysRevB.79.195322},
  url = {https://link.aps.org/doi/10.1103/PhysRevB.79.195322},
  archivePrefix = {arXiv},
  eprint = {cond-mat/0607531},
}

@article{Tanaka2012,
author = {Tanaka ,Yukio and Sato ,Masatoshi and Nagaosa ,Naoto},
title = {Symmetry and Topology in Superconductors –Odd-Frequency Pairing and Edge States–},
journal = {Journal of the Physical Society of Japan},
volume = {81},
number = {1},
pages = {011013},
year = {2012},
doi = {10.1143/JPSJ.81.011013},
URL = { 
        https://doi.org/10.1143/JPSJ.81.011013
},
  archivePrefix = {arXiv},
  eprint = {1105.4700},
  primaryClass = {cond-mat},
}

@article{Fu2011,
  title = {Topological Crystalline Insulators},
  author = {Fu, Liang},
  journal = {Phys. Rev. Lett.},
  volume = {106},
  issue = {10},
  pages = {106802},
  numpages = {4},
  year = {2011},
  month = {3},
  publisher = {American Physical Society},
  doi = {10.1103/PhysRevLett.106.106802},
  url = {https://link.aps.org/doi/10.1103/PhysRevLett.106.106802},
  archivePrefix = {arXiv},
  eprint = {1010.1802},
  primaryClass = {cond-mat},
}

@Article{Hsieh2012,
author={Hsieh, Timothy H.
and Lin, Hsin
and Liu, Junwei
and Duan, Wenhui
and Bansil, Arun
and Fu, Liang},
title={Topological crystalline insulators in the SnTe material class},
journal={Nature Communications},
year={2012},
month={7},
day={31},
volume={3},
number={1},
pages={982},
issn={2041-1723},
doi={10.1038/ncomms1969},
url={https://doi.org/10.1038/ncomms1969},
  archivePrefix = {arXiv},
  eprint = {1202.1003},
  primaryClass = {cond-mat},
}

@Article{Slager2013,
author={Slager, Robert-Jan
and Mesaros, Andrej
and Juri{\v{c}}i{\'{c}}, Vladimir
and Zaanen, Jan},
title={The space group classification of topological band-insulators},
journal={Nature Physics},
year={2013},
month={2},
day={01},
volume={9},
number={2},
pages={98--102},
issn={1745-2481},
doi={10.1038/nphys2513},
url={https://doi.org/10.1038/nphys2513},
  archivePrefix = {arXiv},
  eprint = {1209.2610},
  primaryClass = {cond-mat},
}

@article{Chiu2013,
  title = {Classification of topological insulators and superconductors in the presence of reflection symmetry},
  author = {Chiu, Ching-Kai and Yao, Hong and Ryu, Shinsei},
  journal = {Phys. Rev. B},
  volume = {88},
  issue = {7},
  pages = {075142},
  numpages = {24},
  year = {2013},
  month = {8},
  publisher = {American Physical Society},
  doi = {10.1103/PhysRevB.88.075142},
  url = {https://link.aps.org/doi/10.1103/PhysRevB.88.075142},
  archivePrefix = {arXiv},
  eprint = {1303.1843},
  primaryClass = {cond-mat},
}

@article{Morimoto2013,
  title = {Topological classification with additional symmetries from Clifford algebras},
  author = {Morimoto, Takahiro and Furusaki, Akira},
  journal = {Phys. Rev. B},
  volume = {88},
  issue = {12},
  pages = {125129},
  numpages = {16},
  year = {2013},
  month = {9},
  publisher = {American Physical Society},
  doi = {10.1103/PhysRevB.88.125129},
  url = {https://link.aps.org/doi/10.1103/PhysRevB.88.125129},
  archivePrefix = {arXiv},
  eprint = {1306.2505},
  primaryClass = {cond-mat},
}

@article{Shiozaki2014,
  title = {Topology of crystalline insulators and superconductors},
  author = {Shiozaki, Ken and Sato, Masatoshi},
  journal = {Phys. Rev. B},
  volume = {90},
  issue = {16},
  pages = {165114},
  numpages = {41},
  year = {2014},
  month = {10},
  publisher = {American Physical Society},
  doi = {10.1103/PhysRevB.90.165114},
  url = {https://link.aps.org/doi/10.1103/PhysRevB.90.165114},
  archivePrefix = {arXiv},
  eprint = {1403.3331},
  primaryClass = {cond-mat},
}

@article{Kruthoff2017,
  title = {Topological Classification of Crystalline Insulators through Band Structure Combinatorics},
  author = {Kruthoff, Jorrit and de Boer, Jan and van Wezel, Jasper and Kane, Charles L. and Slager, Robert-Jan},
  journal = {Phys. Rev. X},
  volume = {7},
  issue = {4},
  pages = {041069},
  numpages = {23},
  year = {2017},
  month = {12},
  publisher = {American Physical Society},
  doi = {10.1103/PhysRevX.7.041069},
  url = {https://link.aps.org/doi/10.1103/PhysRevX.7.041069},
  archivePrefix = {arXiv},
  eprint = {1612.02007},
  primaryClass = {cond-mat},
}

@Article{Po2017,
author={Po, Hoi Chun
and Vishwanath, Ashvin
and Watanabe, Haruki},
title={Symmetry-based indicators of band topology in the 230 space groups},
journal={Nature Communications},
year={2017},
month={6},
day={30},
volume={8},
number={1},
pages={50},
issn={2041-1723},
doi={10.1038/s41467-017-00133-2},
url={https://doi.org/10.1038/s41467-017-00133-2},
  archivePrefix = {arXiv},
  eprint = {1703.00911},
  primaryClass = {cond-mat},
}

@Article{Bradlyn2017,
author={Bradlyn, Barry
and Elcoro, L.
and Cano, Jennifer
and Vergniory, M. G.
and Wang, Zhijun
and Felser, C.
and Aroyo, M. I.
and Bernevig, B. Andrei},
title={Topological quantum chemistry},
journal={Nature},
year={2017},
month={7},
day={01},
volume={547},
number={7663},
pages={298--305},
issn={1476-4687},
doi={10.1038/nature23268},
url={https://doi.org/10.1038/nature23268},
  archivePrefix = {arXiv},
  eprint = {1703.02050},
  primaryClass = {cond-mat},
}

@article{Rudner2009,
  title = {Topological Transition in a Non-Hermitian Quantum Walk},
  author = {Rudner, M. S. and Levitov, L. S.},
  journal = {Phys. Rev. Lett.},
  volume = {102},
  issue = {6},
  pages = {065703},
  numpages = {4},
  year = {2009},
  month = {2},
  publisher = {American Physical Society},
  doi = {10.1103/PhysRevLett.102.065703},
  url = {https://link.aps.org/doi/10.1103/PhysRevLett.102.065703},
  archivePrefix = {arXiv},
  eprint = {0807.2048},
  primaryClass = {cond-mat},
}

@article{Sato2012,
    author = {Sato, Masatoshi and Hasebe, Kazuki and Esaki, Kenta and Kohmoto, Mahito},
    title = {Time-Reversal Symmetry in Non-Hermitian Systems},
    journal = {Progress of Theoretical Physics},
    volume = {127},
    number = {6},
    pages = {937--974},
    year = {2012},
    month = {6},
    issn = {0033-068X},
    doi = {10.1143/PTP.127.937},
    url = {https://doi.org/10.1143/PTP.127.937},
  archivePrefix = {arXiv},
  eprint = {1106.1806},
  primaryClass = {cond-mat},
}

@article{Hu2011,
  title = {Absence of topological insulator phases in non-Hermitian $PT$-symmetric Hamiltonians},
  author = {Hu, Yi Chen and Hughes, Taylor L.},
  journal = {Phys. Rev. B},
  volume = {84},
  issue = {15},
  pages = {153101},
  numpages = {4},
  year = {2011},
  month = {10},
  publisher = {American Physical Society},
  doi = {10.1103/PhysRevB.84.153101},
  url = {https://link.aps.org/doi/10.1103/PhysRevB.84.153101},
  archivePrefix = {arXiv},
  eprint = {1107.1064},
  primaryClass = {cond-mat},
}

@article{Esaki2011,
  title = {Edge states and topological phases in non-Hermitian systems},
  author = {Esaki, Kenta and Sato, Masatoshi and Hasebe, Kazuki and Kohmoto, Mahito},
  journal = {Phys. Rev. B},
  volume = {84},
  issue = {20},
  pages = {205128},
  numpages = {19},
  year = {2011},
  month = {11},
  publisher = {American Physical Society},
  doi = {10.1103/PhysRevB.84.205128},
  url = {https://link.aps.org/doi/10.1103/PhysRevB.84.205128},
  archivePrefix = {arXiv},
  eprint = {1107.2079},
  primaryClass = {cond-mat},
}

@article{Schomerus2013,
author = {Henning Schomerus},
journal = {Opt. Lett.},
number = {11},
pages = {1912--1914},
publisher = {Optica Publishing Group},
title = {Topologically protected midgap states in complex photonic lattices},
volume = {38},
month = {6},
year = {2013},
url = {https://opg.optica.org/ol/abstract.cfm?URI=ol-38-11-1912},
doi = {10.1364/OL.38.001912},
  archivePrefix = {arXiv},
  eprint = {1301.0777},
  primaryClass = {physics},
}

@Article{Longhi2015,
author={Longhi, Stefano
and Gatti, Davide
and Valle, Giuseppe Della},
title={Robust light transport in non-Hermitian photonic lattices},
journal={Scientific Reports},
year={2015},
month={8},
day={28},
volume={5},
number={1},
pages={13376},
issn={2045-2322},
doi={10.1038/srep13376},
url={https://doi.org/10.1038/srep13376},
  archivePrefix = {arXiv},
  eprint = {1503.08787},
  primaryClass = {physics},
}

@article{Lee2016,
  title = {Anomalous Edge State in a Non-Hermitian Lattice},
  author = {Lee, Tony E.},
  journal = {Phys. Rev. Lett.},
  volume = {116},
  issue = {13},
  pages = {133903},
  numpages = {5},
  year = {2016},
  month = {4},
  publisher = {American Physical Society},
  doi = {10.1103/PhysRevLett.116.133903},
  url = {https://link.aps.org/doi/10.1103/PhysRevLett.116.133903},
  archivePrefix = {arXiv},
  eprint = {1603.05312},
  primaryClass = {quant-ph},
}

@article{Leykam2017,
  title = {Edge Modes, Degeneracies, and Topological Numbers in Non-Hermitian Systems},
  author = {Leykam, Daniel and Bliokh, Konstantin Y. and Huang, Chunli and Chong, Y. D. and Nori, Franco},
  journal = {Phys. Rev. Lett.},
  volume = {118},
  issue = {4},
  pages = {040401},
  numpages = {6},
  year = {2017},
  month = {1},
  publisher = {American Physical Society},
  doi = {10.1103/PhysRevLett.118.040401},
  url = {https://link.aps.org/doi/10.1103/PhysRevLett.118.040401},
  archivePrefix = {arXiv},
  eprint = {1610.04029},
  primaryClass = {cond-mat},
}

@article{Xu2017,
  title = {Weyl Exceptional Rings in a Three-Dimensional Dissipative Cold Atomic Gas},
  author = {Xu, Yong and Wang, Sheng-Tao and Duan, L.-M.},
  journal = {Phys. Rev. Lett.},
  volume = {118},
  issue = {4},
  pages = {045701},
  numpages = {5},
  year = {2017},
  month = {1},
  publisher = {American Physical Society},
  doi = {10.1103/PhysRevLett.118.045701},
  url = {https://link.aps.org/doi/10.1103/PhysRevLett.118.045701},
  archivePrefix = {arXiv},
  eprint = {1611.02239},
  primaryClass = {cond-mat},
}

@article{Shen2018,
  title = {Topological Band Theory for Non-Hermitian Hamiltonians},
  author = {Shen, Huitao and Zhen, Bo and Fu, Liang},
  journal = {Phys. Rev. Lett.},
  volume = {120},
  issue = {14},
  pages = {146402},
  numpages = {6},
  year = {2018},
  month = {4},
  publisher = {American Physical Society},
  doi = {10.1103/PhysRevLett.120.146402},
  url = {https://link.aps.org/doi/10.1103/PhysRevLett.120.146402},
  archivePrefix = {arXiv},
  eprint = {1706.07435},
  primaryClass = {cond-mat},
}

@article{Kozii2024,
  title = {Non-Hermitian topological theory of finite-lifetime quasiparticles: Prediction of bulk Fermi arc due to exceptional point},
  author = {Kozii, Vladyslav and Fu, Liang},
  journal = {Phys. Rev. B},
  volume = {109},
  issue = {23},
  pages = {235139},
  numpages = {5},
  year = {2024},
  month = {6},
  publisher = {American Physical Society},
  doi = {10.1103/PhysRevB.109.235139},
  url = {https://link.aps.org/doi/10.1103/PhysRevB.109.235139},
  archivePrefix = {arXiv},
  eprint = {1708.05841},
  primaryClass = {cond-mat},
}

@article{Takata2018,
  title = {Photonic Topological Insulating Phase Induced Solely by Gain and Loss},
  author = {Takata, Kenta and Notomi, Masaya},
  journal = {Phys. Rev. Lett.},
  volume = {121},
  issue = {21},
  pages = {213902},
  numpages = {6},
  year = {2018},
  month = {11},
  publisher = {American Physical Society},
  doi = {10.1103/PhysRevLett.121.213902},
  url = {https://link.aps.org/doi/10.1103/PhysRevLett.121.213902},
  archivePrefix = {arXiv},
  eprint = {1710.09169},
  primaryClass = {physics},
}

@article{Gong2018,
  title = {Topological Phases of Non-Hermitian Systems},
  author = {Gong, Zongping and Ashida, Yuto and Kawabata, Kohei and Takasan, Kazuaki and Higashikawa, Sho and Ueda, Masahito},
  journal = {Phys. Rev. X},
  volume = {8},
  issue = {3},
  pages = {031079},
  numpages = {33},
  year = {2018},
  month = {9},
  publisher = {American Physical Society},
  doi = {10.1103/PhysRevX.8.031079},
  url = {https://link.aps.org/doi/10.1103/PhysRevX.8.031079},
  archivePrefix = {arXiv},
  eprint = {1802.07964},
  primaryClass = {cond-mat},
}

@Article{Kawabata2019,
author={Kawabata, Kohei
and Higashikawa, Sho
and Gong, Zongping
and Ashida, Yuto
and Ueda, Masahito},
title={Topological unification of time-reversal and particle-hole symmetries in non-Hermitian physics},
journal={Nature Communications},
year={2019},
month={1},
day={17},
volume={10},
number={1},
pages={297},
issn={2041-1723},
doi={10.1038/s41467-018-08254-y},
url={https://doi.org/10.1038/s41467-018-08254-y},
  archivePrefix = {arXiv},
  eprint = {1804.04676},
  primaryClass = {quant-ph},
}

@article{Yao2018,
  title = {Edge States and Topological Invariants of Non-Hermitian Systems},
  author = {Yao, Shunyu and Wang, Zhong},
  journal = {Phys. Rev. Lett.},
  volume = {121},
  issue = {8},
  pages = {086803},
  numpages = {8},
  year = {2018},
  month = {},
  publisher = {American Physical Society},
  doi = {10.1103/PhysRevLett.121.086803},
  url = {https://link.aps.org/doi/10.1103/PhysRevLett.121.086803},
  archivePrefix = {arXiv},
  eprint = {1803.01876},
  primaryClass = {cond-mat},
}

@article{Yao2018b,
  title = {Non-Hermitian Chern Bands},
  author = {Yao, Shunyu and Song, Fei and Wang, Zhong},
  journal = {Phys. Rev. Lett.},
  volume = {121},
  issue = {13},
  pages = {136802},
  numpages = {8},
  year = {2018},
  month = {9},
  publisher = {American Physical Society},
  doi = {10.1103/PhysRevLett.121.136802},
  url = {https://link.aps.org/doi/10.1103/PhysRevLett.121.136802},
  archivePrefix = {arXiv},
  eprint = {1804.04672},
  primaryClass = {cond-mat},
}

@article{Kunst2018,
  title = {Biorthogonal Bulk-Boundary Correspondence in Non-Hermitian Systems},
  author = {Kunst, Flore K. and Edvardsson, Elisabet and Budich, Jan Carl and Bergholtz, Emil J.},
  journal = {Phys. Rev. Lett.},
  volume = {121},
  issue = {2},
  pages = {026808},
  numpages = {6},
  year = {2018},
  month = {7},
  publisher = {American Physical Society},
  doi = {10.1103/PhysRevLett.121.026808},
  url = {https://link.aps.org/doi/10.1103/PhysRevLett.121.026808},
  archivePrefix = {arXiv},
  eprint = {1805.06492},
  primaryClass = {cond-mat},
}

@article{McDonald2018,
  title = {Phase-Dependent Chiral Transport and Effective Non-Hermitian Dynamics in a Bosonic Kitaev-Majorana Chain},
  author = {McDonald, A. and Pereg-Barnea, T. and Clerk, A. A.},
  journal = {Phys. Rev. X},
  volume = {8},
  issue = {4},
  pages = {041031},
  numpages = {15},
  year = {2018},
  month = {11},
  publisher = {American Physical Society},
  doi = {10.1103/PhysRevX.8.041031},
  url = {https://link.aps.org/doi/10.1103/PhysRevX.8.041031},
  archivePrefix = {arXiv},
  eprint = {1805.12557},
  primaryClass = {cond-mat},
}

@article{Lee2019,
  title = {Anatomy of skin modes and topology in non-Hermitian systems},
  author = {Lee, Ching Hua and Thomale, Ronny},
  journal = {Phys. Rev. B},
  volume = {99},
  issue = {20},
  pages = {201103},
  numpages = {5},
  year = {2019},
  month = {5},
  publisher = {American Physical Society},
  doi = {10.1103/PhysRevB.99.201103},
  url = {https://link.aps.org/doi/10.1103/PhysRevB.99.201103},
  archivePrefix = {arXiv},
  eprint = {1809.02125},
  primaryClass = {cond-mat},
}

@article{Liu2019,
  title = {Second-Order Topological Phases in Non-Hermitian Systems},
  author = {Liu, Tao and Zhang, Yu-Ran and Ai, Qing and Gong, Zongping and Kawabata, Kohei and Ueda, Masahito and Nori, Franco},
  journal = {Phys. Rev. Lett.},
  volume = {122},
  issue = {7},
  pages = {076801},
  numpages = {8},
  year = {2019},
  month = {2},
  publisher = {American Physical Society},
  doi = {10.1103/PhysRevLett.122.076801},
  url = {https://link.aps.org/doi/10.1103/PhysRevLett.122.076801},
  archivePrefix = {arXiv},
  eprint = {1810.04067},
  primaryClass = {cond-mat},
}

@article{Lee2019b,
  title = {Hybrid Higher-Order Skin-Topological Modes in Nonreciprocal Systems},
  author = {Lee, Ching Hua and Li, Linhu and Gong, Jiangbin},
  journal = {Phys. Rev. Lett.},
  volume = {123},
  issue = {1},
  pages = {016805},
  numpages = {6},
  year = {2019},
  month = {7},
  publisher = {American Physical Society},
  doi = {10.1103/PhysRevLett.123.016805},
  url = {https://link.aps.org/doi/10.1103/PhysRevLett.123.016805},
  archivePrefix = {arXiv},
  eprint = {1810.11824},
  primaryClass = {cond-mat},
}

@article{Kawabata2019b,
  title = {Symmetry and Topology in Non-Hermitian Physics},
  author = {Kawabata, Kohei and Shiozaki, Ken and Ueda, Masahito and Sato, Masatoshi},
  journal = {Phys. Rev. X},
  volume = {9},
  issue = {4},
  pages = {041015},
  numpages = {52},
  year = {2019},
  month = {10},
  publisher = {American Physical Society},
  doi = {10.1103/PhysRevX.9.041015},
  url = {https://link.aps.org/doi/10.1103/PhysRevX.9.041015},
  archivePrefix = {arXiv},
  eprint = {1812.09133},
  primaryClass = {cond-mat},
}

@article{Zhou2019,
  title = {Periodic table for topological bands with non-Hermitian symmetries},
  author = {Zhou, Hengyun and Lee, Jong Yeon},
  journal = {Phys. Rev. B},
  volume = {99},
  issue = {23},
  pages = {235112},
  numpages = {19},
  year = {2019},
  month = {6},
  publisher = {American Physical Society},
  doi = {10.1103/PhysRevB.99.235112},
  url = {https://link.aps.org/doi/10.1103/PhysRevB.99.235112},
  archivePrefix = {arXiv},
  eprint = {1812.10490},
  primaryClass = {cond-mat},
}

@article{Herviou2019,
  title = {Defining a bulk-edge correspondence for non-Hermitian Hamiltonians via singular-value decomposition},
  author = {Herviou, Lo\"{\i}c and Bardarson, Jens H. and Regnault, Nicolas},
  journal = {Phys. Rev. A},
  volume = {99},
  issue = {5},
  pages = {052118},
  numpages = {10},
  year = {2019},
  month = {5},
  publisher = {American Physical Society},
  doi = {10.1103/PhysRevA.99.052118},
  url = {https://link.aps.org/doi/10.1103/PhysRevA.99.052118},
  archivePrefix = {arXiv},
  eprint = {1901.00010},
  primaryClass = {cond-mat},
}

@article{Zirnstein2021,
  title = {Bulk-Boundary Correspondence for Non-Hermitian Hamiltonians via Green Functions},
  author = {Zirnstein, Heinrich-Gregor and Refael, Gil and Rosenow, Bernd},
  journal = {Phys. Rev. Lett.},
  volume = {126},
  issue = {21},
  pages = {216407},
  numpages = {7},
  year = {2021},
  month = {5},
  publisher = {American Physical Society},
  doi = {10.1103/PhysRevLett.126.216407},
  url = {https://link.aps.org/doi/10.1103/PhysRevLett.126.216407},
  archivePrefix = {arXiv},
  eprint = {1901.11241},
  primaryClass = {cond-mat},
}

@article{Borgnia2020,
  title = {Non-Hermitian Boundary Modes and Topology},
  author = {Borgnia, Dan S. and Kruchkov, Alex Jura and Slager, Robert-Jan},
  journal = {Phys. Rev. Lett.},
  volume = {124},
  issue = {5},
  pages = {056802},
  numpages = {6},
  year = {2020},
  month = {2},
  publisher = {American Physical Society},
  doi = {10.1103/PhysRevLett.124.056802},
  url = {https://link.aps.org/doi/10.1103/PhysRevLett.124.056802},
  archivePrefix = {arXiv},
  eprint = {1902.07217},
  primaryClass = {cond-mat},
}

@article{Kawabata2019c,
  title = {Classification of Exceptional Points and Non-Hermitian Topological Semimetals},
  author = {Kawabata, Kohei and Bessho, Takumi and Sato, Masatoshi},
  journal = {Phys. Rev. Lett.},
  volume = {123},
  issue = {6},
  pages = {066405},
  numpages = {7},
  year = {2019},
  month = {8},
  publisher = {American Physical Society},
  doi = {10.1103/PhysRevLett.123.066405},
  url = {https://link.aps.org/doi/10.1103/PhysRevLett.123.066405},
  archivePrefix = {arXiv},
  eprint = {1902.08479},
  primaryClass = {cond-mat},
}

@article{Yokomizo2019,
  title = {Non-Bloch Band Theory of Non-Hermitian Systems},
  author = {Yokomizo, Kazuki and Murakami, Shuichi},
  journal = {Phys. Rev. Lett.},
  volume = {123},
  issue = {6},
  pages = {066404},
  numpages = {6},
  year = {2019},
  month = {8},
  publisher = {American Physical Society},
  doi = {10.1103/PhysRevLett.123.066404},
  url = {https://link.aps.org/doi/10.1103/PhysRevLett.123.066404},
  archivePrefix = {arXiv},
  eprint = {1902.10958},
  primaryClass = {cond-mat},
}

@article{Okuma2019,
  title = {Topological Phase Transition Driven by Infinitesimal Instability: Majorana Fermions in Non-Hermitian Spintronics},
  author = {Okuma, Nobuyuki and Sato, Masatoshi},
  journal = {Phys. Rev. Lett.},
  volume = {123},
  issue = {9},
  pages = {097701},
  numpages = {6},
  year = {2019},
  month = {8},
  publisher = {American Physical Society},
  doi = {10.1103/PhysRevLett.123.097701},
  url = {https://link.aps.org/doi/10.1103/PhysRevLett.123.097701},
  archivePrefix = {arXiv},
  eprint = {1904.06355},
  primaryClass = {cond-mat},
}

@article{Lee2019c,
  title = {Topological Correspondence between Hermitian and Non-Hermitian Systems: Anomalous Dynamics},
  author = {Lee, Jong Yeon and Ahn, Junyeong and Zhou, Hengyun and Vishwanath, Ashvin},
  journal = {Phys. Rev. Lett.},
  volume = {123},
  issue = {20},
  pages = {206404},
  numpages = {6},
  year = {2019},
  month = {Nov},
  publisher = {American Physical Society},
  doi = {10.1103/PhysRevLett.123.206404},
  url = {https://link.aps.org/doi/10.1103/PhysRevLett.123.206404},
  archivePrefix = {arXiv},
  eprint = {1906.08782},
  primaryClass = {cond-mat},
}

@Article{Wanjura2020,
author={Wanjura, Clara C.
and Brunelli, Matteo
and Nunnenkamp, Andreas},
title={Topological framework for directional amplification in driven-dissipative cavity arrays},
journal={Nature Communications},
year={2020},
month={6},
day={19},
volume={11},
number={1},
pages={3149},
issn={2041-1723},
doi={10.1038/s41467-020-16863-9},
url={https://doi.org/10.1038/s41467-020-16863-9},
  archivePrefix = {arXiv},
  eprint = {1909.11647},
  primaryClass = {cond-mat},
}

@article{Zhang2020,
  title = {Correspondence between Winding Numbers and Skin Modes in Non-Hermitian Systems},
  author = {Zhang, Kai and Yang, Zhesen and Fang, Chen},
  journal = {Phys. Rev. Lett.},
  volume = {125},
  issue = {12},
  pages = {126402},
  numpages = {6},
  year = {2020},
  month = {9},
  publisher = {American Physical Society},
  doi = {10.1103/PhysRevLett.125.126402},
  url = {https://link.aps.org/doi/10.1103/PhysRevLett.125.126402},
  archivePrefix = {arXiv},
  eprint = {1910.01131},
  primaryClass = {cond-mat},
}

@article{Okuma2020,
  title = {Topological Origin of Non-Hermitian Skin Effects},
  author = {Okuma, Nobuyuki and Kawabata, Kohei and Shiozaki, Ken and Sato, Masatoshi},
  journal = {Phys. Rev. Lett.},
  volume = {124},
  issue = {8},
  pages = {086801},
  numpages = {7},
  year = {2020},
  month = {2},
  publisher = {American Physical Society},
  doi = {10.1103/PhysRevLett.124.086801},
  url = {https://link.aps.org/doi/10.1103/PhysRevLett.124.086801},
  archivePrefix = {arXiv},
  eprint = {1910.02878},
  primaryClass = {cond-mat},
}
\let\addcontentsline\oldaddcontentsline

\clearpage

\section*{End Matter}

\textit{Notation}.---
The Hamiltonians introduced in the proof of bulk-boundary correspondence are summarized as follows.
\begin{align*}
    \begin{array}{l|l}
        \text{PBC}\quad (\boldsymbol{k}\in\text{BZ}) & \text{SIBC}\quad (X=\mathbb{Z}_{\ge 0}\times\mathbb{Z}^{d-1}) \\ \hline
        H(\boldsymbol{k})\curvearrowright\mathcal{H} & H^{\SIBC}\curvearrowright\mathcal{H}\otimes\ell^2(X) \\
        H(\boldsymbol{k})|_{\mathcal{M}}\colon\mathcal{M}\to\mathcal{M}' & H|_{\mathcal{M}}^{\SIBC}\colon\mathcal{M}\otimes\ell^2(X)\to\mathcal{M}'\otimes\ell^2(X) \\
        \tilde{H}(\boldsymbol{k})\curvearrowright\mathcal{M}\oplus\mathcal{M}' & \tilde{H}^{\SIBC}\curvearrowright(\mathcal{M}\oplus\mathcal{M}')\otimes\ell^2(X)
    \end{array}
\end{align*}
Here we use a symbol $A\curvearrowright\mathcal{K}$ to denote that $A$ is an operator mapping a space $\mathcal{K}$ to itself, $A\colon\mathcal{K}\to\mathcal{K}$

\textit{Subspace-protected topological phase in even spatial dimension and quadratic edge dispersion}.---
In the main text, we have defined the subspace-protected winding number \eqref{eq:winding} in odd dimension.
Yet, the notion of subspace-protected topological phases can be extended into even dimensions by imposing a symmetry constraint on the restricted Hamiltonian.

For example, let us consider a class of Bloch Hamiltonians that satisfy the subspace property with $\dim\mathcal{M}=\dim\mathcal{M}'$ and whose restricted Hamiltonians $H(\boldsymbol{k})|_\mathcal{M}\colon\mathcal{M}\to\mathcal{M}'$ have a chiral symmetry:
\begin{align}\label{eq:chiral_symmetry}
    \Gamma^\dagger H(\boldsymbol{k})|_\mathcal{M}^\dagger
    = -H(\boldsymbol{k})|_\mathcal{M}\Gamma
\end{align}
with a unitary operator $\Gamma\colon\mathcal{M}'\to\mathcal{M}$.
Note that $\mathrm{i}H(\boldsymbol{k})\Gamma$ is Hermitian.
Introducing a Green function by 
\begin{align}
    G_\mathcal{M}(\omega,\boldsymbol{k})
    \coloneqq (\mathrm{i}\omega - \mathrm{i}H(\boldsymbol{k})|_\mathcal{M}\Gamma)^{-1} \curvearrowright\mathcal{M}',\quad 
    \omega\in\mathbb{R}
\end{align}
we can construct an integer-valued topological invariant protected by both the subspace property and the chiral symmetry for even dimension $d=2n$ as \cite{Kawabata2019b}
\begin{align}\label{eq:Chern_number}
    C^\mathcal{M}
    \coloneqq \frac{n!}{(2\pi\mathrm{i})^{n+1}(2n+1)!}\int_{\mathbb{R}_\omega\times\text{BZ}} \operatorname{Tr}_\mathcal{M}\left[ (G_\mathcal{M}\mathrm{d}G_\mathcal{M}^{-1})^{2n+1} \right],
\end{align}
which is nothing but the $n$-th Chern number for the Hermitian matrix $\mathrm{i}H(\boldsymbol{k})\Gamma$.

A concrete model in two spatial dimensions is given by
\begin{gather}
    \label{eq:AIII_model}
    H(k_x,k_y)
    = \begin{pmatrix}
        0 & H_1(k_x,k_y)^\dagger \\
        H_1(k_x,k_y) & H_2(k_x,k_y)
    \end{pmatrix}, \\
    \begin{split}
        H_1(k_x,k_y)
        &= - \mathrm{i} s\sigma_x + (m-\cos k_x-\cos k_y)\sigma_y \\
        &\qquad- \sin k_y \sigma_z -\mathrm{i}\sin k_x,
    \end{split} \\
    \begin{split}
        H_2(k_x,k_y)
        &= d_0+d_{0,x}\sin k_x+d_{0,y}\sin k_y \\
        &+ \sum_{i=x,y,z}(d_i+d_{i,x}\sin k_x+d_{i,y}\sin k_y)\sigma_i
    \end{split}
\end{gather}
with $m,s\in\mathbb{R}$, $d_\mu,d_{\mu,x},d_{\mu,y}\in\mathbb{R}$ ($\mu=0,x,y,z$).
The whole Hermitian Hamiltonian \eqref{eq:AIII_model} has no symmetry due to the presence of $H_2(k_x,k_y)$ but satisfies the subspace property with $\mathcal{M}'=\mathcal{M}^\perp=0\oplus0\oplus\mathbb{C}\oplus\mathbb{C}$.
The restricted Hamiltonian $H(k_x,k_y)|_\mathcal{M}=H_1(k_x,k_y)$ is non-Hermitian and has the chiral symmetry \eqref{eq:chiral_symmetry} with $\Gamma=\sigma_x$.
The Hermitian matrix $\mathrm{i}H_1(k_x,k_y)\sigma_x$ is identical with the energy-shifted Qi--Wu--Zhang model \cite{Qi2006}, 
\begin{align}
    \sin k_x \sigma_x + \sin k_y \sigma_y + (m-\cos k_x-\cos k_y)\sigma_z + s,
\end{align}
so its Chern number \eqref{eq:Chern_number} is given by
\begin{align}
    C^\mathcal{M}
    = \begin{dcases*}
        0 & for $m<-2$ \& $|s|<|m|-2$ \\
        1 & for $-2<m<0$ \& $|s|<\min\{|m|,2-|m|\}$ \\
        -1 & for $0<m<2$ \& $|s|<\min\{m,2-m\}$ \\
        0 & for $m>2$ \& $|s|<m-2$.
    \end{dcases*}
\end{align}

\begin{figure}
    \centering
    \includegraphics[width=\linewidth]{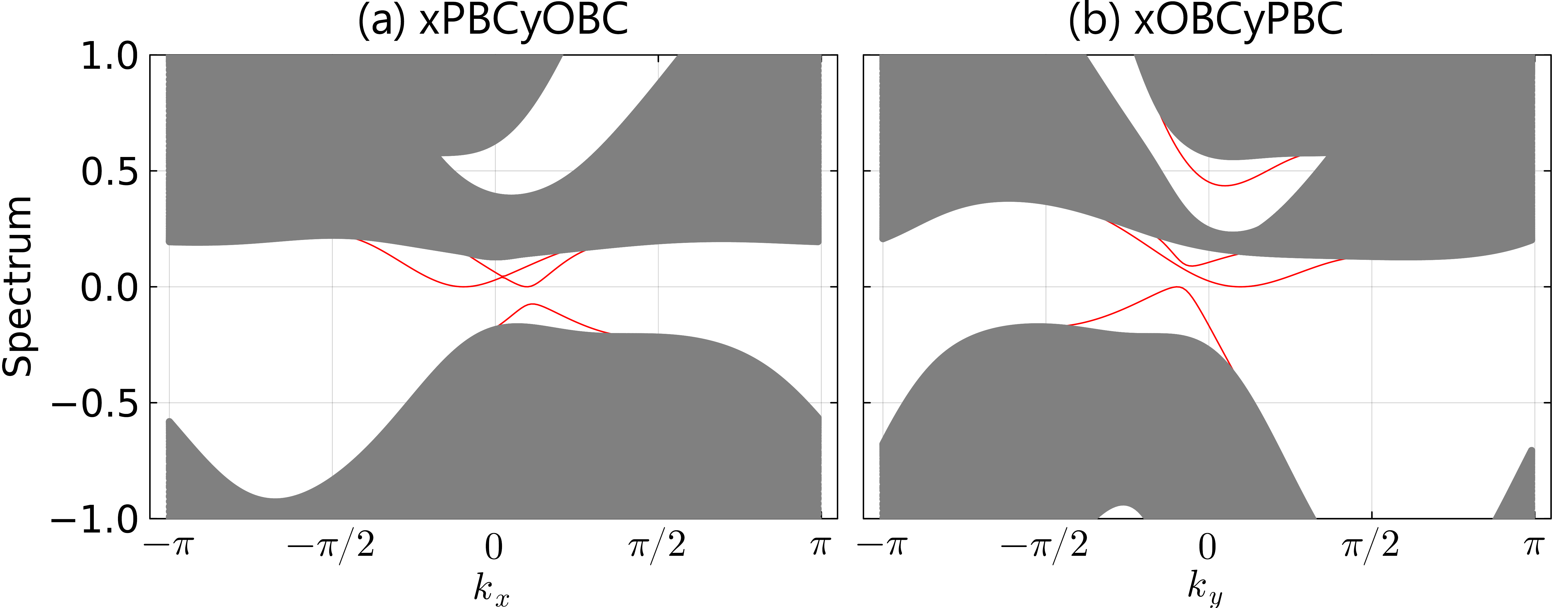}
    \caption{Bulk and edge energy spectra of the Hermitian model \eqref{eq:AIII_model}.
    The parameters are choses as $m=1$, $s=0.3$, $d_0=-1.871$, $d_x=-0.878$, $d_y=1.706$, $d_z=-0.364$, $d_{0,x}=0.214$, $d_{x,x}=-1.215$, $d_{y,x}=1.412$, $d_{z,x}=-0.369$, $d_{0,y}=-1.165$, $d_{x,y}=-0.529$, $d_{y,y}=-1.342$, and $d_{z,y}=0.676$.
    The gray regions are bulk spectra obtained from the Bloch Hamiltonian $H(k_x,k_y)$ for both (a) and (b).
    The red lines are edge spectra under xPBC \& yOBC for (a) and under xOBC \& yPBC for (b).
    We take the system size along the OBC direction as $L=30$ when calculating the edge spectra from the real-space Hamiltonian.
    }
    \label{fig:2D_chiral}
\end{figure}

In Fig.~\ref{fig:2D_chiral}, we show the bulk and edge energy dispersion at $m=1$ and $s=0.3$ belonging to the nontrivial phase with $C^\mathcal{M}=-1$.
We can see the edge spectra (red lines) touching $E=0$ at two points in $k$-space, which manifests the bulk-boundary correspondence in the present case.
The two topological zero modes are localized on opposite edges, being consistent with $|C^{\mathcal{M}}|=1$.

Importantly, the edge spectra do not flow from the lower bulk bands to the upper bulk bands.
This is because the subspace property with $\mathcal{M}\neq\mathcal{M}'$ protects not the spectral flow but only zero energy modes due to instability of the subspace property against energy shift.
For the same reason, a similar phenomenon, the absence of spectral flow of boundary modes, occurs in Hermitian topological insulators and superconductors whose topology is protected only by either sublattice or particle-hole symmetry \cite{Altland2024,Nakamura2024b,Shiozaki2024}.

It is also noteworthy that the edge spectra have a quadratic or higher-order energy dispersion around $E=0$.
This property is stable within the present class due to subspace-protected zero-energy pinning of the edge spectra.
Among conventional topological matter, such a quadratic dispersion of boundary modes typically appears in topological crystalline insulators \cite{Fu2011} because a certain crystalline symmetry prohibits any linear terms of $\boldsymbol{k}$ in the boundary effective Hamiltonian.
On the other hand, our model \eqref{eq:AIII_model} does not need any crystalline symmetry to stabilize the quadratic edge dispersion.
This is a characteristic of subspace-protected topological phases in higher dimension.

\textit{Subspace property with $\mathcal{M}=\mathcal{M}'$ or $\mathcal{M}\perp\mathcal{M}'$}.---
Here, we supplement the explanation of the guideline for the search of physical platforms possessing the subspace property by focusing on the matrix form of the Hamiltonian.

If $\mathcal{M}=\mathcal{M}'$, the subspace property $H(\boldsymbol{k})\mathcal{M}\subseteq\mathcal{M}$ implies $\mathcal{M}$ is an invariant subspace of $H(\boldsymbol{k})$.
In the orthonormal basis of $\mathcal{H}$ as $\ket{e_1},\ldots,\ket{e_m}\in\mathcal{M}$, $\ket{e_{m+1}},\ldots,\ket{e_n}\in\mathcal{M}^\perp$ with $n=\dim\mathcal{H}$, the Hamiltonian $H(\boldsymbol{k})$ takes a block-triangular form, i.e., $H(\boldsymbol{k}) = \begin{pmatrix} H(\boldsymbol{k})|_\mathcal{M} & * \\ 0 & * \end{pmatrix}.$
Such a triangular structure appears in the self-energy of a fermionic Green function in the Keldysh formalism, $\Sigma = \begin{pmatrix} \Sigma^{\mathrm{R}} & \Sigma^{\mathrm{K}} \\ 0 & \Sigma^{\mathrm{A}} \end{pmatrix}$, where $\Sigma^{\mathrm{R}}$, $\Sigma^{\mathrm{A}}$, and $\Sigma^{\mathrm{K}}$ denote the retarded, advanced, and Keldysh components, respectively.
Therefore, the subspace property is a ubiquitous structure in non-equilibrium fermionic systems.

Next, if $\mathcal{M}\perp\mathcal{M}'$, the Hamiltonian in the basis above is written as $H(\boldsymbol{k}) = \begin{pmatrix} 0 & * \\ H(\boldsymbol{k})|_\mathcal{M} & * \end{pmatrix}.$
Imposing the Hermiticity provides 
\begin{gather}
    H(\boldsymbol{k}) 
    = H_{\text{off-diag}}(\boldsymbol{k}) + H_{\text{diag}}(\boldsymbol{k}), \\
    H_{\text{off-diag}}(\boldsymbol{k})
    = \begin{pmatrix} 
        0 & H(\boldsymbol{k})|_\mathcal{M}^\dagger \\ 
        H(\boldsymbol{k})|_\mathcal{M} & 0 
    \end{pmatrix},\ 
    H_{\text{diag}}(\boldsymbol{k})
    = \begin{pmatrix}
        0 & 0 \\
        0 & V(\boldsymbol{k})
    \end{pmatrix},
\end{gather}
where $V(\boldsymbol{k})\curvearrowright\mathcal{M}^\perp$ is a Hermitian matrix. 
This form of Hamiltonian admits two types of interpretation.

First, we can regard $H_{\text{off-diag}}$ as the original system and $H_{\text{diag}}$ as a perturbation.
From this point of view, the original system has the sublattice symmetry but the perturbation $H_{\text{diag}}$ breaks it.
For example, the Bernal-stacked bilayer graphene possesses the sublattice symmetry if only nearest-neighbor hoppings are considered. 
The system consists of four atoms in the unit cell, labeled by A1, B1 on the upper layer and A2, B2 on the lower layer.
The $4\times 4$ Hamiltonian is of the form
\begin{align}
    H_{\text{off-diag}}
    = \begin{pmatrix} 
        0 & 0 & -\gamma_0 f^* & 0 \\ 
        0 & 0 & \gamma_1 & -\gamma_0 f^* \\ 
        -\gamma_0 f & \gamma_1 & 0 & 0 \\ 
        0 & -\gamma_0 f & 0 & 0 
    \end{pmatrix} 
    \begin{array}{l} 
        \mathrm{A}1 \\ 
        \mathrm{A}2 \\ 
        \mathrm{B}1 \\ 
        \mathrm{B}2, 
    \end{array}
\end{align}
where $\gamma_0$ is the intralayer coupling between A and B atoms, $f$ is a $\boldsymbol{k}$-dependent function given by $f(\boldsymbol{k})=\sum_{\ell=1}^3 e^{\mathrm{i}\boldsymbol{k}\cdot\boldsymbol{\delta}_\ell}$ with $\boldsymbol{\delta}_1=(0,1/\sqrt{3})$, $\boldsymbol{\delta}_2=(1/2,-1/(2\sqrt{3}))$, $\boldsymbol{\delta}_3=(-1/2,-1/(2\sqrt{3}))$, and $\gamma_1$ is the interlayer coupling between B1 and A2 atoms.
However, the boron nitride (BN) substrate gives an effective onsite staggered potential $\Delta$ to the lower layer, which yields the perturbed term $H_{\text{diag}}=\operatorname{diag}(0,\Delta,0,-\Delta)$ breaking the sublattice symmetry.
Still, the total Hamiltonian $H=H_{\text{off-diag}}+H_{\text{diag}}$ has the form 
\begin{align}
    H
    = \begin{pmatrix} 
        0 & 0 & -\gamma_0 f^* & 0 \\ 
        0 & \Delta & \gamma_1 & -\gamma_0 f^* \\ 
        -\gamma_0 f & \gamma_1 & 0 & 0 \\ 
        0 & -\gamma_0 f & 0 & -\Delta 
    \end{pmatrix} 
    \begin{array}{l} 
        \mathrm{A}1 \\ 
        \mathrm{A}2 \\ 
        \mathrm{B}1 \\ 
        \mathrm{B}2, 
    \end{array}
\end{align}
and hence possesses the subspace property $H(\boldsymbol{k})\ket{\mathrm{A}1}=-\gamma_0f(\boldsymbol{k})\ket{\mathrm{B}1}$.
Therefore, the bilayer graphene on the BN substrate is a promising platform of the subspace property.

Second, we can identify $H_{\text{diag}}$ as the original system and $H_{\text{off-diag}}$ as a perturbation. 
From this perspective, $H_{\text{diag}}$ is interpreted as a decoupled system of dispersive bands and flat bands, and $H_{\text{off-diag}}$ as a hybridization of both bands.
This is an ideal situation of heavy fermion systems consisting of itinerant and localized electrons if electron-electron interactions are ignored or renormalized to a free-fermion system.
To see this, we write the effective Hamiltonian $H_{\text{eff}}(\boldsymbol{k},\omega)=H_0(\boldsymbol{k})+\Sigma(\boldsymbol{k},\omega)$ extracted from the single-particle Green function of a heavy fermion system as $H_{\text{eff}} = \begin{pmatrix} \varepsilon_{f} & V^* \\ V & \varepsilon_{c} \end{pmatrix}$, where $\varepsilon_c,\varepsilon_f$ are the dispersions of itinerant and localized electrons, respectively, and $V$ is the hybridization.
Here, $H_{\text{eff}}$ is generally non-Hermitian due to a finite quasiparticle lifetime.
If $\varepsilon_f$ is sufficiently small compared to other terms, $H_{\text{eff}}$ can be approximated by $H_{\text{approx}} = \begin{pmatrix} 0 & V^* \\ V & \varepsilon_c \end{pmatrix}$, which has the subspace property with $\mathcal{M}\perp\mathcal{M}'$.
Lastly we remark that the bulk subspace-protected topological phase is no longer well-defined for $H_{\text{eff}}$ due to the presence of $\varepsilon_f$, but a boundary mode can lie in the gap even for $H_{\text{eff}}$ if $H_{\text{approx}}$ hosts a subspace-protected boundary zero mode.
This is because we expect in the absence of non-Hermitian skin effects that a small perturbation yields a small variation of eigenvalues.

\textit{$\mathbb{Z}_2$ invariant for the stacked model \eqref{eq:stacked_model}}.---
The stacked model \eqref{eq:stacked_model} of two copies of the same Hamiltonian \eqref{eq:BDI_model} in the nontrivial $\mathbb{Z}_2$ phase should be in the trivial phase.
We can check this by direct calculation of the $\mathbb{Z}_2$ topological invariant.
By using the PHS operator $C=\operatorname{diag}(\sigma_x,\sigma_x)$ for $H_{\text{stack}}(k)|_{\mathcal{M}\oplus\mathcal{M}}$, the subspace-protected $\mathbb{Z}_2$ invariant for the stacked model is indeed calculated as
\begin{align}
    (-1)^\nu
    &= \sgn\left[ \frac{\operatorname{Pf}(H_{\text{stack}}(\pi)|_{\mathcal{M}\oplus\mathcal{M}}C)}{\operatorname{Pf}(H_{\text{stack}}(0)|_{\mathcal{M}\oplus\mathcal{M}}C)} \right] \notag \\
    &= \sgn\left[ \frac{(h-t)^2+c^2}{(h+t)^2+c^2} \right]
    = 1,
\end{align}
yielding $\nu=0$.
Reflecting this bulk behavior, a boundary zero mode should disappear in the stacked model, as shown in Fig.~\ref{fig:BDI}.

\end{document}